\documentclass[aps,pra,amssymb,floatfix]{revtex4}
\usepackage{bm}
\usepackage{amsmath}
\usepackage{enumerate}
\usepackage{hyperref}
\usepackage{subfigure}
\usepackage{graphicx}
\usepackage{epstopdf} 

\begin{document}
\title{General Features of the Relaxation Dynamics of Interacting Quantum Systems}

\author{E. J. Torres-Herrera}
\author{Manan Vyas}
\author{Lea F. Santos}
\affiliation{Department of Physics, Yeshiva University, New York, New York 10016, USA}

\date{\today}

\begin{abstract}
We study numerically and analytically isolated interacting quantum systems that  are taken out of equilibrium instantaneously (quenched). The probability of finding the initial state in time, the so-called fidelity, decays fastest for systems described by full random matrices, where simultaneous many-body interactions are implied. In the realm of realistic systems with two-body interactions, the dynamics is slower and depends on the interplay between the initial state and the Hamiltonian characterizing the system. The fastest fidelity decay in this case is Gaussian and can persist until saturation. A simple general picture, in which the fidelity plays a central role, is also achieved for the short-time dynamics of few-body observables. It holds for initial states that are eigenstates of the observables. We also discuss the need to reassess analytical expressions that were previously proposed to describe the evolution of the Shannon entropy. Our analyses are mainly developed for initial states that can be prepared in experiments with  cold atoms in optical lattices.
\end{abstract}


\maketitle

\section{Introduction}

Despite the ubiquity of many-body quantum systems out of equilibrium, they are much less understood than quantum systems in equilibrium. To advance our understanding and to construct a general picture, it is necessary to identify the elements that lead to similar dynamics. Determining how fast these systems evolve in time~\cite{Bhattacharyya1983,Pfeifer1993,GiovannettiPRA2003,Giovannetti2003,Giovannetti2004} is also essential for the development of algorithms for quantum optimal control~\cite{Caneva2009}. In these two contexts, the unitary evolution of isolated many-body quantum systems is of particular interest due, in part, to the connection with current experiments in optical lattices~\cite{Greiner2002,Kinoshita06,hofferberth07,Bloch2008,Trotzky2008,Cazalilla2011,Chen2011,Aidelsburger2011,Simon2011,Trotzky2012,Fukuhara2013}.  The latter are quasi-isolated systems, where coherent evolutions can be studied for very long times.

The evolution of an isolated system can be initiated by changing instantaneously the parameters of a certain initial Hamiltonian which is brought into a new final Hamiltonian. This abrupt perturbation is referred to as a quench. The system starts off in an eigenstate of the  initial Hamiltonian. The fidelity (return probability)~\cite{Gorin2006,LoschRef}, which is defined as the overlap between the initial state and its evolved counterpart, is a way to characterize the system evolution. This quantity is related to the Loschmidt echo. It is also analogous to the characteristic function of the probability distribution of work~\cite{Silva2008,Silva2009,Gambassi2011,Heyl2012} and is therefore likely to find applications in quantum thermodynamics, particularly in studies related with the quantification of the work done to take quantum systems out of equilibrium. The fidelity decays exponentially when the final Hamiltonian is chaotic~\cite{Peres1984,expRef,Cerruti2002,FlambaumARXIV,Flambaum2000A,Flambaum2001a,Flambaum2001b,Weinstein2003,Izrailev2006}. In fact, this behavior is expected to hold even in integrable Hamiltonians provided the initial state be sufficiently delocalized in the energy eigenbasis~\cite{Emerson2002,Santos2012PRL,Santos2012PRE}.

Here, we extend the results obtained in Ref.~\cite{Torres2014PRA89} and show that the fidelity can have a faster than exponential behavior. The fidelity corresponds to the Fourier transform of the energy distribution of the initial state. This distribution is referred to here as local density of states (LDOS)  for any initial state~\cite{Flambaum2000}. In the case of realistic final Hamiltonians with two-body interactions, the maximum LDOS is Gaussian. In this scenario, the fidelity decay is therefore also Gaussian~\cite{Flambaum2001a,Flambaum2001b,Izrailev2006,Torres2014PRA89} and this behavior can persist until saturation~\cite{Torres2014PRA89}. The slower exponential decay observed in previous studies occurs when the energy distribution of the initial state is restricted to a Breit-Wigner (Lorentzian) form or to a Gaussian shape that is not well filled. There are, however, situations where even the Gaussian decay can be surpassed. One, addressed here, happens when the system is described by full random matrices. This is not a very realistic approach, since full random matrices imply simultaneous interactions of many particles, but it serves to identify the lower bound for the fidelity decay in many-body quantum systems, where the initial state has a single-peaked energy distribution. 

Another essential aspect of nonequilibrium dynamics, especially in connection with experiments, is the evolution of few-body observables. A complete description is a complex enterprise, since the evolution depends not only on the initial state and final Hamiltonian, but also on the individual properties of the various observables.  However, when the initial state is also an eigenstate of the observables, their dynamics depends explicitly on the results for the fidelity and a simple general picture becomes available. In this case, the short-time dynamics is quadratic in time. We find distinct observables evolving according to very different Hamiltonians, but showing a very similar behavior.

We also discuss results for the evolution of the Shannon entropy. The fidelity gives the probability of finding the initial state in time, whereas the Shannon entropy captures the participation of other states. Analytical expressions were obtained showing that the Shannon entropy increases linearly in time in the limit of strong perturbation~\cite{Flambaum2001b,Santos2012PRL,Santos2012PRE}. Even though this behavior is reproduced for the initial states considered here,  the results do not match those previous analytical expressions. We speculate on the causes for the discrepancy and how it may be solved.

The core sections of this paper are Secs.~\ref{Sec:ldos}, \ref{Sec:ini}, and \ref{Sec:Obs}.
Section \ref{Sec:ldos} is the central one; it contains the main results about the relationship between fidelity decay and  energy distribution of the initial state. 
Section \ref{Sec:ini} extends this discussion to initial states that are accessible to experiments with cold atoms in optical lattices. 
Section \ref{Sec:Obs} analyzes the short-time dynamics of few-body observables. In the other sections, we cover the description of the model, their density of states, and the meaning of quench dynamics [Sec.~\ref{Sec:dos}], as well as the results for the evolution of the Shannon entropy [Sec.~\ref{Sec:Shannon}].
Concluding remarks are presented in Sec.~\ref{Sec:Summary}.

\section{Model, Density of States, and Quench}
\label{Sec:dos}

A way to describe many-body quantum systems is to treat them statistically using  full random matrices. This was Wigner's approach to describe heavy nuclei~\cite{Wigner1951}  and it was soon employed in the description of other complex systems, such as atoms, molecules, and quantum dots~\cite{HaakeBook,ReichlBook,StockmannBook,Guhr1998,Gubin2012}. However, full random matrices do not capture the details of realistic quantum systems with few-body interactions, as the spin-1/2 systems considered here. Below we give a general overview of the differences between the two.

\subsection{Full Random Matrices}

Full random matrices are matrices filled with random numbers. Their only constraint is to satisfy the symmetries of the system they are trying to describe. The distribution of the spacings $s$ between neighboring energy levels has a Wigner-Dyson shape, $\Pi_{ WD}(s)$, indicating level repulsion. The exact shape of $\Pi_{ WD}(s)$ depends on the symmetries of the system. Ensembles of real and symmetric random matrices, the so-called Gaussian Orthogonal Ensembles (GOE's), imply time reversal invariance and lead to $\Pi_{ WD}(s) = (\pi s/2)\exp(-\pi s^2/4)$. Level repulsion is one of the main features of what is called quantum chaos~\cite{HaakeBook,ReichlBook,StockmannBook,Guhr1998,Gubin2012}. The latter corresponds to properties of eigenvalues and eigenstates found in the quantum level that indicate whether the system in the classical level is chaotic or not. The term has in fact been extended to refer to those properties even in quantum systems without a classical limit.

\begin{figure}[htb]
\centering
\includegraphics*[width=4.25in]{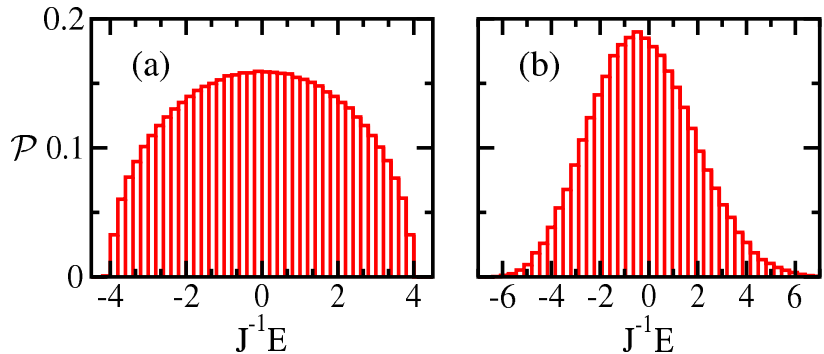}
\caption{(Color online) Density of states for a full random matrix from a GOE  (a) and for $\widehat{H}_{\Delta=0.5,\lambda=1}$ in the ${\cal \widehat{S}}^z=0$ subspace of $L=16$ (b). The GOE is normalized so that the length of the spectrum is $8J$ and its width is similar to that of $\widehat{H}_{\Delta=0.5,\lambda=1}$ (cf. Eq.~\ref{deltaEfrm} and Table~\ref{table:DOS});  ${\cal D} = 12\,870$.}
\label{fig:density}
\end{figure}

The density of states ${\cal P}(E)$ (not to be confused with the local density of states $P(E)$ analyzed in Sec.~\ref{Sec:ldos}) of full random matrices has a semicircular shape~\cite{Wigner1957,Wigner1967,Guhr1998}. 
This is shown in Fig.~\ref{fig:density} (a) for a full random matrix from a GOE.

The problem with full random matrices is that they imply the existence of interactions that can change the states of many particles at once. In contrast, real systems involve few-body interactions, usually just two-body interactions. Early attempts to improve this picture led to the introduction of band random matrices~\cite{Wigner1955} and two-body random ensembles~\cite{French1970,Bohigas1971,Brody1981,Flores2001}. Similar to the latter, the systems that we consider have only two-body interactions, but they do not involve any randomness.

\subsection{Spin-1/2 Model}

The realistic spin-1/2 models that we investigate here  describe real magnetic compounds~\cite{Sologubenko2000PRB,Hess2007,Hlubek2010}, crystals of fluorapatite~\cite{Cappellaro2007,Cappellaro2007b,Ramanathan2011}, and have also been simulated with optical lattices~\cite{Trotzky2008,Simon2011,Fukuhara2013}. We focus on one-dimensional lattice systems with open boundaries and an even number $L$ of sites. The Hamiltonian contains nearest-neighbor (NN) and possibly also next-nearest-neighbor (NNN) couplings,
\begin{eqnarray}
&& \widehat{H} = \widehat{H}_{NN} + \lambda \; \widehat{H}_{NNN} \;;
\label{ham} \\
&& \widehat{H}_{NN} = \sum_{j=1}^{L-1} J \left(\widehat{S}_j^x \widehat{S}_{j+1}^x + \widehat{S}_j^y \widehat{S}_{j+1}^y +\Delta \widehat{S}_j^z \widehat{S}_{j+1}^z \right) \;,
\nonumber \\
&& \widehat{H}_{NNN} = \sum_{j=1}^{L-2} J \left(\widehat{S}_j^x \widehat{S}_{j+2}^x + \widehat{S}_j^y \widehat{S}_{j+2}^y +\Delta \widehat{S}_j^z \widehat{S}_{j+2}^z \right) \;.
\nonumber 
\end{eqnarray}
It can be mapped onto systems of spinless fermions~\cite{Jordan1928} or hardcore bosons~\cite{Holstein1940}. In the equation above,  $\hbar=1$ and $\widehat{S}^{x,y,z}_j=\widehat{\sigma}^{x,y,z}/2$ are the spin operators at site $j$; $\widehat{\sigma}_j^{x,y,z}$ being the Pauli spin matrices. The coupling strength $J$, the anisotropy parameter $\Delta$, and the ratio $\lambda$ between NNN and NN exchanges are chosen positive, thus favoring antiferromagnetic order. $\widehat{S}_j^x \widehat{S}_{j+1}^x + \widehat{S}_j^y \widehat{S}_{j+1}^y$ $(\widehat{S}_j^x \widehat{S}_{j+2}^x + \widehat{S}_j^y \widehat{S}_{j+2}^y)$ is the flip-flop term and $\widehat{S}_j^z \widehat{S}_{j+1}^z (\widehat{S}_j^z \widehat{S}_{j+2}^z)$ is the Ising interaction between NN (NNN) spins.  

The Hamiltonian conserves total spin in the $z$ direction, $[\widehat{H}, \;{\cal \widehat{S}}^z]=0$, where ${\cal \widehat{S}}^z = \sum_{j=1}^L \widehat{S}_j^z$. Other symmetries include parity, invariance under a global $\pi$ rotation around the $x$ axis when  ${\cal \widehat{S}}^z=0$, and conservation of total spin  ${\cal \widehat{S}}^2=(\sum_{j=1}^L \vec{S}_j)^2$ when $\Delta=1$. We work in the  ${\cal \widehat{S}}^z=0$ subspace, where the dimension of the Hamiltonian matrix is  ${\cal D}=\binom{L}{L/2}$.

The noninteracting XX model ($\Delta = \lambda = 0$) is trivially solved. The interacting XXZ case ($\Delta \neq 0$, $\lambda =0$) is solved with the Bethe ansatz  \cite{Bethe1931}. The system undergoes a crossover to the chaotic regime as $\lambda $ increases \cite{Gubin2012, Santos2012PRE, Zangara2013,Kudo2005}, the level spacing distribution gradually changing from a Poisson distribution, $\Pi_{ P}(s) = \exp(-s)$, in the case of the XXZ model~\cite{noteXX}, to the GOE Wigner-Dyson form.

We analyze the dynamics of the system for the following choices of parameters for the final Hamiltonian, 
\begin{center}
\begin{tabular}{ll}
  \hline
Integrable isotropic & $\widehat{H}_{\Delta=1,\lambda=0}$      \\
\hline
Integrable anisotropic &$\widehat{H}_{\Delta=0.5,\lambda=0}$     \\
\hline
Weakly chaotic isotropic & $\widehat{H}_{\Delta=1,\lambda=0.4}$       \\
\hline
Weakly chaotic anisotropic \hspace{0.7 cm}  & $\widehat{H}_{\Delta=0.5,\lambda=0.4}$         \\
\hline
Strongly chaotic isotropic & $\widehat{H}_{\Delta=1,\lambda=1}$     \\
\hline
Strongly chaotic anisotropic & $\widehat{H}_{\Delta=0.5,\lambda=1}$        \\
\hline
\end{tabular}
\end{center}
Note that the value of $\lambda$ leading to chaos depends on the system size. The larger the system, the smaller the parameter needs to be. In the thermodynamic limit, an infinitesimally small perturbation may be enough to break the integrability of the system~\cite{Santos2010PRE}.

The density of states of the above Hamiltonians, independent of the regime of the system, is Gaussian, as shown in Fig.~\ref{fig:density} (b) for $\widehat{H}_{\Delta=0.5,\lambda=1}$ (see other illustrations in \cite{Zangara2013}). This is typical of systems with two-body interactions \cite{French1970,Brody1981,ZelevinskyRep1996, Izrailev1990, Kota2001,Manan2010}. The majority of the states are close to the middle of the spectrum, where strong mixing occurs. Thus, the eigenstates reach their highest level of delocalization in the center of the spectrum and are more localized close to the edges.

\begin{table}[h]
\caption{Width and center of the Gaussian fit for the density of states; $L=16$; ${\cal S}^z=0$.}
\begin{center}
\begin{tabular}{|l|c|c|}
\hline 
  &  \hspace{0.3 cm} $\omega $ \hspace{0.3 cm} &  \hspace{0.2 cm} $\langle E \rangle$  \hspace{0.2 cm}  \\
  \hline
$\widehat{H}_{\Delta=1,\lambda=0}$ &  1.76  &   -0.119    \\
\hline
$\widehat{H}_{\Delta=0.5,\lambda=0}$ &  1.53  &  -0.039   \\
\hline
$\widehat{H}_{\Delta=1,\lambda=0.4}$ &  1.87  & -0.368      \\
\hline
$\widehat{H}_{\Delta=0.5,\lambda=0.4}$ &   1.64 & -0.106        \\
\hline
$\widehat{H}_{\Delta=1,\lambda=1}$ &  2.40 &   -0.571    \\
\hline
$\widehat{H}_{\Delta=0.5,\lambda=1}$ &  2.11  &  -0.356     \\
\hline
\end{tabular}
\end{center}
\label{table:DOS}
\end{table}
The width $\omega $ and the average energy $\langle E \rangle $ obtained from a Gaussian fit for the density of states of the Hamiltonians above are shown in Table~\ref{table:DOS}. The distributions get broader as the value of the anisotropy parameter and the strength of NNN couplings increase. They also shift their center away from zero and become more assymetric.

\subsection{Quench Dynamics}
\label{Sec:quench}

The scenario we consider here is that of a quench. The initial state  $|\Psi(0)\rangle = |\text{ini}\rangle$ is as an eigenstate of an initial (unperturbed) Hamiltonian $\widehat{H}_\text{I}$. The dynamics starts with the sudden change of some parameter(s) of this Hamiltonian in a time interval much shorter than any characteristic time scale of the model. This results in the final (perturbed) Hamiltonian $\widehat{H}_\text{F}=\widehat{H}_\text{I}+\widehat{V}$ with eigenvalues $E_\alpha$ and eigenstates $|\psi_\alpha\rangle\neq| \text{ini} \rangle$; $\widehat{V}$ being the perturbation.  The unitary time evolution  is given by
\begin{equation}
|\Psi(t)\rangle=e^{-i\widehat{H}_\text{F}t}| \text{ini} \rangle=\sum_{\alpha} C_{\alpha}^{\text{ini}}  e^{-iE_\alpha t}|\psi_\alpha\rangle ,
\label{eq:instate}
\end{equation} 
where the coefficients $C_{\alpha}^{\text{ini}} = \langle \psi_{\alpha} | \text{ini} \rangle $  are the overlaps of the initial state with the eigenstates of $\widehat{H}_\text{F}$. 

The evolution is computed numerically with full exact diagonalization for  $L\leq 16$ (${\cal D} = 12\,870$) and with EXPOKIT~\cite{Expokit,Sidje1998} for larger system sizes. We examine up to $L=24$ (${\cal D} = 2\,704\,156$). 
EXPOKIT is a software package based on Krylov subspace projection methods.
Instead of diagonalizing the complete system Hamiltonian, the package
 computes directly the action of the matrix exponential $e^{-i\widehat{H}_\text{F}t}$ on a vector
of interest.


\section{LDOS and Fidelity}
\label{Sec:ldos}

The fidelity is one of our main quantities of interest. It corresponds to the probability of finding the system still in the initial state after time $t$. It is given by the overlap,
\begin{equation}
F(t) \equiv  |\langle \Psi(0)| \Psi(t) \rangle |^2 =\left| \langle \text{ini} | e^{-i \widehat{H}_\text{F} t} |\text{ini} \rangle \right|^2 
= \left|\sum_{\alpha} |C_{\alpha}^{\text{ini}} |^2 e^{-i E_{\alpha} t}  \right|^2 .
\label{eq:fidelity}
\end{equation}
$F(t)$ is therefore equivalent to the Fourier transform in energy of the components $|C_{\alpha}^{\text{ini}}|^2$.
The distribution $P^{\text{ini}}(E) = \sum_\alpha |C_\alpha^\text{ini}|^2\delta(E-E_\alpha)$ of the components $|C_{\alpha}^{\text{ini}}|^2$ in the eigenvalues $E_{\alpha}$ is referred to here as LDOS. We obtain the LDOS numerically by dividing the whole range of the spectrum of  $\widehat{H}_F$ in small windows of energy and computing the sum $\sum_{\alpha} |C_{\alpha}^{\text{ini}}|^2$ inside each bin. Experimental measures of the density of states and LDOS is a subject of intense investigation, particularly in nuclear physics~\cite{ZelevinskyRep1996,Egidy2005,Burger2012,Isaak2013}. From the connection between LDOS and the probability distribution of work, we can infer also the possibility of measuring the LDOS experimentally with Ramsey interferometric techniques~\cite{Goold2011,Dorner2013}.

At very short times, 
\begin{equation}
F(t)\approx  \left| e^{-i E_\text{ini} t}\left[\sum_{\alpha} |C_{\alpha}^{\text{ini}} |^2-i\sum_{\alpha} |C_{\alpha}^{\text{ini}} |^2(E_\alpha-E_\text{ini})t-\frac{1}{2}\sum_{\alpha} |C_{\alpha}^{\text{ini}} |^2(E_\alpha-E_\text{ini})^2t^2\right] \right|^2  \approx 1-\sigma_\text{ini}^2t^2,
\label{eq:fidelity_approx}
\end{equation}
where
\begin{equation}
\sigma_{\text{ini}} = \sqrt{\sum_{\alpha} |C_{\alpha}^{\text{ini}} |^2 (E_{\alpha} - E_{\text{ini}})^2}
=\sqrt{\sum_{n \neq \text{ini}} |\langle n |\widehat{H}_F | \text{ini}\rangle |^2 },
\label{deltaE}
\end{equation}
is the uncertainty in energy and 
\begin{equation}
E_{\text{ini}} = \langle \text{ini} |\widehat{H}_F | \text{ini} \rangle = \sum_{\alpha} |C_{\alpha}^{\text{ini}}|^2 E_{\alpha} 
\label{Eini}
\end{equation}
is the energy of the initial state projected on the final Hamiltonian. In Eq.~(\ref{deltaE}), $|n\rangle$ corresponds to the eigenstates of $\widehat{H}_\text{I}$ and the basis in which the final Hamiltonian is written. Thus, $\sigma_{\text{ini}} $ depends only on the sum of the square of the off-diagonal elements of  $\widehat{H}_\text{F}$ and can be obtained before the diagonalization of this Hamiltonian. 

The approximation in Eq.~(\ref{eq:fidelity_approx}) is valid for any initial state and final Hamiltonian. 
Below, we analyze the fidelity decay for longer times and specific shapes of the distribution $P^\text{ini}(E)$. We also substitute the sum in $\alpha$ by an integral, which is appropriate when ${\cal D}$ is large.

\subsection{Semicircular LDOS}
\label{sec:SC}

\begin{figure}[htb]
\centering
\includegraphics*[width=4.in]{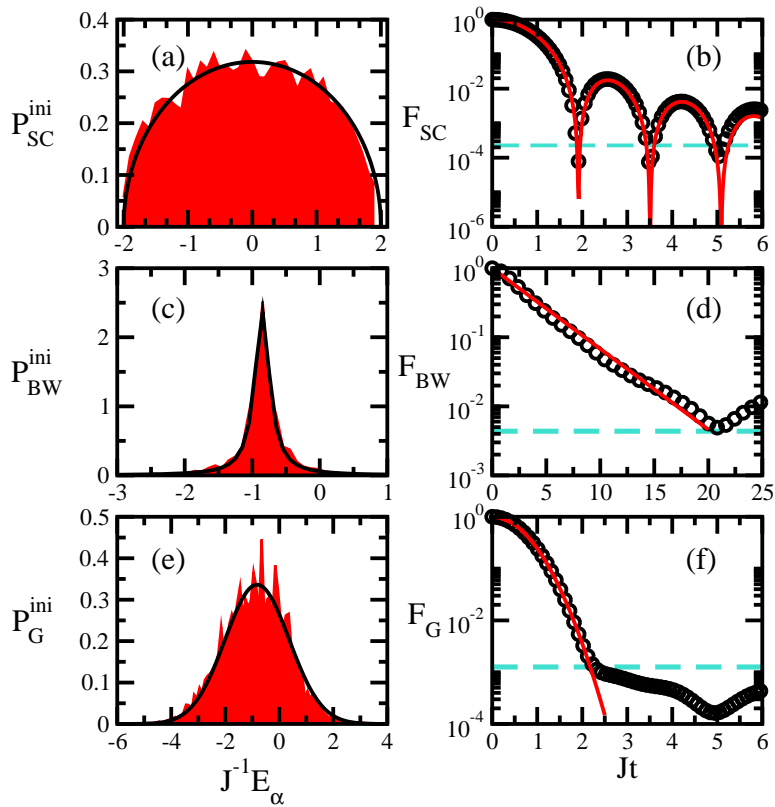}
\caption{(Color online) 
Local density of states (left) and fidelity decay (right). The initial state is an eigenstate of a full random matrix from a GOE; it is projected onto another full random matrix from a GOE.  The random numbers are normalized so that the length of the spectrum is $4J$ and $\sigma_{\text{ini}}= J$ (a,b).
The initial state is an eigenstate from $\widehat{H}_{\Delta=0.5,\lambda=0}$; it is projected onto $\widehat{H}_{\Delta=0.5,\lambda=0.4}$, $E_{\text{ini}} = - 0.803 J^{-1}$ [equivalently, temperature $k_BT = 3.9 J^{-1}$ from $E_{\text{ini}}=\sum_\alpha E_\alpha e^{(-E_\alpha/k_B T)}/\sum_\alpha e^{(-E_\alpha/k_B T)}$, where $k_B$ is Boltzmann constant] and $\Gamma_{\text{ini}}=0.27 J$ (c,d). The initial state is an eigenstate  from $\widehat{H}_{\Delta=0.5,\lambda=0}$; it is projected onto $\widehat{H}_{\Delta=0.5,\lambda=1}$, $E_{\text{ini}}=-0.821 J^{-1}$ [$k_BT=7.1 J^{-1}$] and $\sigma_{\text{ini}}=1.19 J$ (e,f). Solid lines give the analytical expressions, shaded area (left) and circles (right) are numerical results. The saturation value of the fidelity is indicated with the dashed horizontal line; ${\cal \widehat{S}}^z=0$,  $L=16$, ${\cal D} = 12\,870$.}
\label{fig:LDOS}
\end{figure}

For an initial state projected onto a full random matrix, $P^{\text{ini}}(E)$ agrees with the density of states and has again the semicircular shape. The envelope of the distribution is the function
\begin{equation}
P^{\text{ini}}_{SC}(E)= \frac{2}{\pi {\cal E}} \sqrt{1 - \left(\frac{E}{{\cal E}}\right)^2},
\end{equation}
where $2{\cal E}$ is the length of the spectrum and
\begin{eqnarray}
\sigma_{\text{ini}} = \sqrt{ \int_{-{\cal E}}^{{\cal E}} P^{\text{ini}}_{SC}(E) E^2 dE} = \frac{{\cal E} }{2}.
\label{deltaEfrm}
\end{eqnarray}
An illustration is provided in Fig.~\ref{fig:LDOS} (a). 
 
The fidelity for this distribution is given by
\begin{equation}
F_{\text{SC}}(t) = \left| \int_{-{\cal E}}^{{\cal E}} P^{\text{ini}}_{SC}(E) e^{-i E t} dE
\right|^2
= \frac{ [{\cal J}_1( 2 \sigma_{\text{ini}} t)]^2}{\sigma_{\text{ini}}^2 t^2}
\label{Ffrm}
\end{equation}
where  ${\cal J}_1$ is the Bessel function of the first kind. The behavior of the fidelity is in excellent agreement with the numerical results, as shown in Fig.~\ref{fig:LDOS} (b)

The maximum fidelity decay, when the LDOS has a single energy peak, is therefore given by Eq.~(\ref{Ffrm}). Apart from very short times (\ref{eq:fidelity_approx}), even Eq.~(\ref{Ffrm}) is slower than the bound, $F(t)\geq \cos^2(\sigma_{\text{ini}} t)$, derived from the time-energy uncertainty relation~\cite{Bhattacharyya1983,Pfeifer1993,GiovannettiPRA2003,Giovannetti2003,Giovannetti2004}. This latter result can be approached when the energy distribution of $|\text{ini} \rangle$ involves more peaks well separated in energy. This is beyond the scope of this work. Here, we focus on the general scenario of quenches, where the energy distribution of the initial states is usually single peaked.

As mention before, in realistic systems with few-body interactions, the density of states is Gaussian instead of semicircular. This has consequences to the LDOS, which cannot therefore exceed the Gaussian shape.

\subsection{Breit-Wigner  LDOS}

In systems with two-(few-)body interactions and $E_{\text{ini}}$ close to the middle of the spectrum, as the strength of the instantaneous perturbation applied on $\widehat{H}_\text{I}$ increases from zero, LDOS broadens from a delta function to a Breit-Wigner form delineated by ~\cite{ZelevinskyRep1996,Flambaum1997,Flambaum2000,Santos2012PRL,Santos2012PRE},
\begin{equation}
P^{\text{ini}}_{BW}(E) = \frac{1}{2\pi} \frac{\Gamma_{\text{ini}} }{(E_{\text{ini}} - E)^2 +
 \Gamma_{\text{ini}}^2 /4},
\label{BW_SF}
\end{equation}
where $\Gamma_{\text{ini}} $ is the width of the distribution.
An example is given in Fig.~\ref{fig:LDOS} (c), where the quench considered is from the initial Hamiltonian $\widehat{H}_{\Delta=0.5, \lambda=0}$  to the final weakly chaotic Hamiltonian $\widehat{H}_{\Delta=0.5, \lambda=0.4}$. The energy $E_{\text{ini}}$ of the initial state used is away from the edges of the spectrum.

The above distribution leads to the exponential fidelity decay~\cite{expRef,Cerruti2002,FlambaumARXIV,Flambaum2000A,Flambaum2001a,Flambaum2001b,Weinstein2003,Emerson2002,Izrailev2006},
\begin{equation}
F_{\text{BW}}(t) = \left| \int_{-\infty}^{\infty} P^{\text{ini}}_{BW}(E) e^{-i E t} dE
\right|^2
= e^{- \Gamma_{\text{ini}} t},
\label{Fbw}
\end{equation}
as illustrated in Fig.~\ref{fig:LDOS} (d). The quadratic behavior (\ref{eq:fidelity_approx}) of very short times soon switches to an exponential decay.

\subsection{Gaussian  LDOS}

As the perturbation to $\widehat{H}_\text{I}$ increases even further and we eventually reach the regime of strong perturbation, the LDOS of initial states with $E_{\text{ini}}$ away from the edges of the spectrum approaches a Gaussian form~\cite{ZelevinskyRep1996,Frazier1996,Flambaum1997,Flambaum2000,Flambaum2001a,Flambaum2001b,Santos2012PRL,Santos2012PRE} of width (\ref{deltaE}).

In systems with two-body interactions, the Gaussian envelope of the LDOS,
\begin{equation}
P^{\text{ini}}_{G}(E) = \frac{1}{ \sqrt{ 2 \pi \sigma^2_{\text{ini}} }  } \exp \left[ -\frac{(E-E_{\text{ini}})^2}{ 2 \sigma^2_{\text{ini}} }   \right],
\label{Gauss_SF}
\end{equation}
gives the maximum possible spreading of the initial state in the eigenvalues of the final Hamiltonian and is known as the energy shell~\cite{ZelevinskyRep1996,Flambaum1997,Flambaum2000,Flambaum2000A,Flambaum2001a,Flambaum2001b,Santos2012PRL,Santos2012PRE,Torres2013}. The ergodic filling of the energy shell is used as a definition of chaotic states. When this happens, the components $|C_{\alpha}^{\text{ini}} |^2$ of the state become random numbers following the Gaussian distribution. In this sense, a chaotic state may emerge even when one of the Hamiltonians involved is integrable.

The Gaussian distribution leads to a Gaussian fidelity decay controlled by $\sigma_{\text{ini}} $,
\begin{equation}
F_{\text{G}}(t) = \left| \int_{-\infty}^{\infty} P^{\text{ini}}_{G}(E) e^{-i E t} dE
\right|^2
= e^{- \sigma_{\text{ini}}^2  t^2}.
\label{eq:Fgauss}
\end{equation}
This Gaussian behavior was derived also via the central limit theorem~\cite{Venuti2010}.
Illustrations for $P^{\text{ini}}_{G}(E) $ and $F_{\text{G}}(t)$ are given in Figs.~\ref{fig:LDOS} (e) and (f), respectively. The initial state chosen  is an eigenstate of $\widehat{H}_{\Delta=0.5, \lambda=0}$ and it evolves according to the final Hamiltonian $\widehat{H}_{\Delta=0.5, \lambda=1}$. Its energy is away from the edges of the spectrum.

We stress that the fidelity decay can be Gaussian {\em until saturation}, as seen in Fig.~\ref{fig:LDOS} (f) (the saturation point is indicated with the dashed horizontal lines). This is in contrast with previous works, where despite the Gaussian LDOS, the expectation was for an initial Gaussian decay, switching to exponential before saturation~\cite{FlambaumARXIV,Flambaum2000A,Flambaum2001a,Flambaum2001b,Izrailev2006,Santos2012PRL,Santos2012PRE}. In addition to the quench in Fig.~\ref{fig:LDOS} (f) and the ones studied in Sec.~\ref{Sec:FidBasis}, we found various other examples of the Gaussian behavior until saturation, including quenches involving the XX model and spin-1/2 systems with impurities. The fidelity decay is invariably Gaussian for initial states from full random matrices projected into final Hamiltonians with two-body interactions.

As $E_{\text{ini}}$ approaches the border of the spectrum, the initial state becomes more localized, the energy shell less filled, and the LDOS acquires a skewed Gaussian shape~\cite{Flambaum1994,Torres2013}. This is  a consequence of the low density of states at the edges of the spectrum. Examples of this dependence on energy are provided in Figs.~\ref{fig:shellPS} and \ref{fig:shellNS} in Sec.~\ref{Sec:ini}.

\subsection{Relaxation Time}

Equilibration in isolated quantum systems happens in a probabilistic sense. After a long time, for a system without too many degeneracies and with a large Hilbert space, the observables simply fluctuate around their infinite time averages. The size of these fluctuations decreases exponentially with $L$ \cite{Zangara2013}. 

In the particular case of the fidelity,
\[
F(t) = \sum_{\alpha} |C_{\alpha}^{\text{ini}} |^4 + \sum_{\alpha \neq \beta} |C_{\alpha}^{\text{ini}} |^2 |C_{\beta}^{\text{ini}} |^2 e^{i (E_{\alpha} - E_{\beta}) t} ,
\]
the averaging out of the off-diagonal terms at $t \rightarrow \infty$ leads to the infinite time average,
\[
\overline{F}=\sum_{\alpha} |C_{\alpha}^{\text{ini}} |^4=\text{IPR}_{\text{ini}}^{-1} ,
\]
where the inverse participation ratio, $\text{IPR}_{\text{ini}} $, measures the level of delocalization of the initial state in the energy eigenbasis. A large value indicates a delocalized state. 

We define the relaxation time, $t_R$, as the time it takes for the fidelity to first reach the saturation value $\overline{F}$, after which it fluctuates around the average. The variance of the temporal fluctuations is given by
\begin{equation}
\sigma^2_F = \overline{ |F(t)  - \overline{ F(t) }|^2}  
= 
\mathop{\sum_{\alpha \neq \beta} }_{\gamma \neq \delta} 
|C_{\alpha}^{\text{ini}}|^2  |C_{\beta}^{\text{ini}}|^2 |C_{\gamma}^{\text{ini}}|^2  |C_{\delta}^{\text{ini}}|^2
\overline{e^{i (E_{\alpha} - E_{\beta}+E_{\gamma} - E_{\delta}) t}} 
= \text{IPR}_{\text{ini}}^{-2 } + \sum_{\alpha} 
|C_{\alpha}^{\text{ini}}|^8 ,
\label{Eq:sigmaF}  
\end{equation}
where the last equality is obtained for $E_{\alpha} - E_{\beta} = E_{\delta} - E_{\gamma}$.

The saturation point is minimum when $|\text{ini} \rangle $ is quenched to a full random matrix (or when the initial state is extracted from a full random matrix). The eigenstates of full random matrices are random vectors, therefore $|C_{\alpha}^{\text{ini}} |^2$ is on average equal to $1/{\cal D}$, and for GOEs, $\text{IPR}_{\text{ini}} \sim {\cal D}/3$ \cite{Izrailev1990,ZelevinskyRep1996}. 
The relaxation time for an initial state evolved according to full random matrices can thus be obtained from
\begin{equation}
\frac{ [{\cal J}_1( 2 \sigma_{\text{ini}} t_R)]^2}{\sigma_{\text{ini}}^2 t_R^2} = \frac{3}{{\cal D}} .
\end{equation}

In the case where both $\widehat{H}_\text{I}$ and $\widehat{H}_\text{F}$ are two-body-interaction Hamiltonians, in addition to the strength of the perturbation, the level of delocalization of the initial state depends on its energy.  When $E_{\text{ini}}$ is close to the middle of the spectrum, $\text{IPR}_{\text{ini}}$ is large, although usually smaller than ${\cal D}/3$, and it gets smaller as $E_{\text{ini}}$ approaches the borders. The minimum relaxation time for systems with two-body interactions and a single-peaked LDOS is therefore,
\begin{equation}
\exp \left( - \sigma_{\text{ini}}^2  t_R^2 \right) =\text{IPR}_{\text{ini}}^{-1} \Rightarrow t_R=\frac{\sqrt{\ln(\text{IPR}_{\text{ini}}) }}{\sigma_{\text{ini}} }
\end{equation}
The width and filling of the energy shell determine the lifetime of $|{\text{ini}}\rangle$.


\section{Experimentally Accessible Initial States}
\label{Sec:ini}

In this section we extend the studies about fidelity decay to initial states that can be prepared experimentally with cold atoms in optical lattices. They are states where each lattice site has a spin either pointing up or pointing down  in the $z$ direction~\cite{Zangara2013,Santos2011,Pozsgay_2013,PozsgayARXIV}:
\begin{center}
\begin{tabular}{ll}
  \hline
Sharp domain wall &  $|\rm{DW}\rangle = | \uparrow \uparrow \uparrow \ldots \downarrow \downarrow \downarrow \rangle$    \\
\hline
Pairs of parallel spins \hspace{0.6 cm} &  $|\rm{PS}\rangle=| \downarrow \uparrow \uparrow  \downarrow  \downarrow \uparrow \uparrow  \ldots \rangle$   \\
\hline
N\'eel state & $|\rm{NS}\rangle= |\downarrow \uparrow \downarrow \uparrow  \ldots  \downarrow \uparrow \downarrow  \uparrow \rangle$    \\
\hline
\end{tabular}
\end{center}
The proposals for the preparation of domain walls in optical lattices require the application of a magnetic field gradient~\cite{Weld2009}.
The N\'eel state~\cite{Trotzky2008,Koetsier2008,Mathy2012} is similar to the state prepared in~\cite{Trotzky2012} , where only even sites were initially populated and the evolutions of quasi-local densities, currents, and coherences were experimentally investigated after the quench.

We recall that in the quench dynamics considered here, the initial state is an eigenstate of $\widehat{H}_\text{I}$. The eigenstates of the initial Hamiltonian also define the basis in which $\widehat{H}_\text{F}$ is written.
For initial states where each excitation is confined to a single site, as above,  $\widehat{H}_\text{I}$ corresponds to the Ising interaction of Hamiltonian (\ref{ham}). We refer to these states as site-basis vectors (they are also often called computational basis or natural basis). In this basis, the diagonal elements of the final Hamiltonian matrix  depend on $\Delta $ and $\lambda$, while the off-diagonal elements depend only on $\lambda$, since they follow from the flip-flop terms.  

We study how the initial states above evolve according to the final Hamiltonians of Table~\ref{table:DOS}. This corresponds to a nonperturbative quench, where the off-diagonal elements of $\widehat{H}_\text{F}$ are much larger than the
average level spacing. The shape of the energy distributions of these initial states is close to Gaussian, although the energy shell is not always well filled. Better fillings are associated with $E_{\text{ini}}$ closer to the middle of the spectrum. We discuss these distributions in detail in the next subsection before presenting the results for the fidelity.

For site-basis vectors, it is straightforward to calculate  analytically $\langle n |\widehat{H}_F | \text{ini}\rangle$ and, from it, the center $E_{\text{ini}}$ and the width $\sigma_{\text{ini}}$ of the energy shell. One sees that Eq.~(\ref{deltaE}) reduces to
\begin{equation}
\sigma_{\text{ini}} = \frac{J}{2} \sqrt{M_1 + \lambda^2   M_2 } ,
\label{deltaEm}
\end{equation} 
where the connectivity $M_1$ ($M_2$) corresponds to the number of states directly coupled to $|\text{ini}\rangle$ via the NN (NNN) flip-flop term.  The values of  $E_{\text{ini}}$ and $\sigma_{\text{ini}}$ for the three states above are given in Table~\ref{table:initial}.
\begin{table}[h]
\caption{Energy of $|\text{ini}\rangle$ and width of its energy distribution.}
\begin{center}
\begin{tabular}{ccc}
\hline 
\hline 
  &    $E_{\text{ini}}$  & $\sigma_{\text{ini}}$ \\ [0.1 cm]
$|\rm{DW}\rangle$   & \hspace{0.2 cm} $\frac{\displaystyle J\Delta}{\displaystyle 4} [(L-3) + (L-6)\lambda]$ & $ \frac{\displaystyle J}{\displaystyle 2} \sqrt{1+2\lambda^2}$    \\ [0.2 cm] 
 
$ |\rm{PS}\rangle$   & $ - \frac{\displaystyle J\Delta}{\displaystyle 4}  [ 1 + (L-2)\lambda ]$ &  $\frac{\displaystyle J}{\displaystyle 2} \sqrt{ \frac{\displaystyle L}{\displaystyle 2} + (L-2)\lambda^2}$ \\  [0.2 cm]

$ |\rm{NS}\rangle$  & \hspace{0.2 cm} $ 
 \frac{\displaystyle J\Delta}{\displaystyle 4}  [ -(L-1) + (L-2)\lambda ]$ &   $\frac{\displaystyle J}{\displaystyle 2} \sqrt{L-1}$  \\  [0.2 cm]
\hline
\hline 
\end{tabular}
\end{center}
\label{table:initial}
\end{table}

Notice that the total connectivity $M$ of any site-basis vector is low, $M\!=\!M_1 + M_2 \propto L \ll {\cal D}$. However, the eigenstates of the final Hamiltonians in this basis, $|\psi_{\alpha}\rangle = \sum_n C_{\alpha}^n |n\rangle $, can be very delocalized. As a result, the initial state can be very spread out in the energy eigenbasis and the energy shell can therefore be well filled.

\subsection{Gaussian LDOS}

\begin{figure}[htb]
\centering
\includegraphics*[width=4.in]{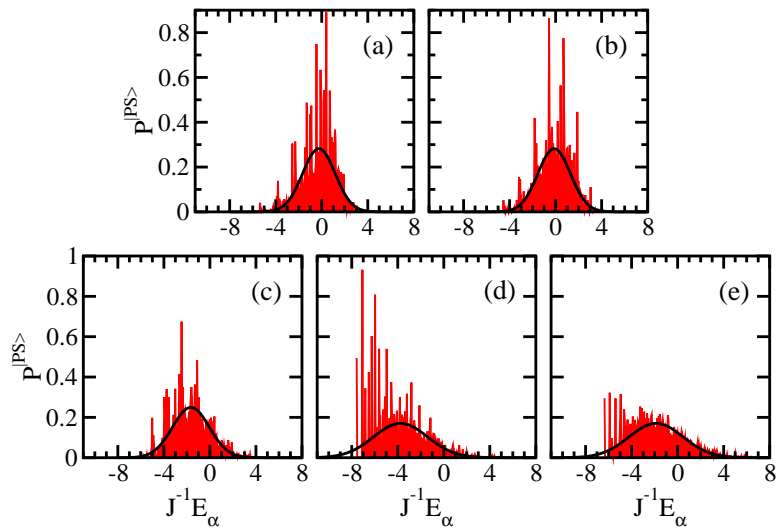}
\caption{(Color online) LDOS for $|\rm{PS}\rangle$. The final Hamiltonians are: (a) $\widehat{H}_{\Delta=1,\lambda=0}$; (b) $\widehat{H}_{\Delta=0.5,\lambda=0}$; (c) $\widehat{H}_{\Delta=1,\lambda=0.4}$; (d) $\widehat{H}_{\Delta=1,\lambda=1}$ and (e) $\widehat{H}_{\Delta=0.5,\lambda=1}$. The solid line is the energy shell: Gaussian centered at $E_{\text{ini}}$ of width $\sigma_{\text{ini}}$ (see Tables \ref{table:initial}, \ref{table:ipr}); bin size = 0.05 $J$; $L=16$.}
\label{fig:shellPS}
\end{figure}

\begin{figure}[htb]
\centering
\includegraphics*[width=4.in]{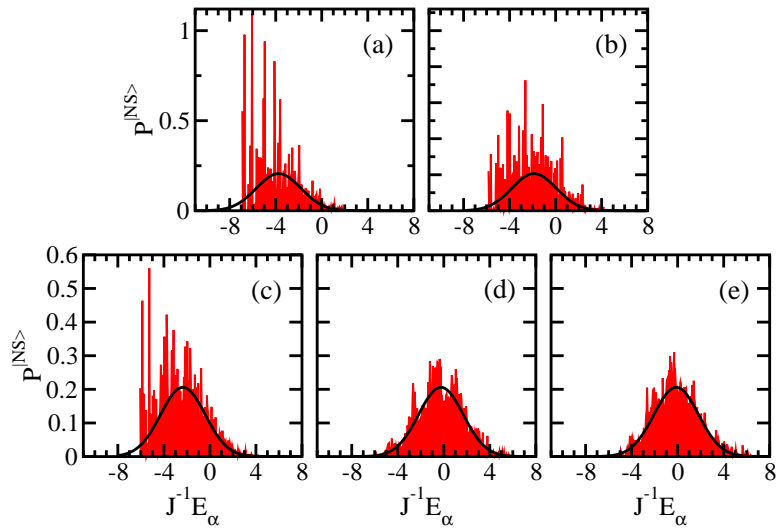}
\caption{(Color online) LDOS for $|\rm{NS}\rangle$. The final Hamiltonians are: (a) $\widehat{H}_{\Delta=1,\lambda=0}$; (b) $\widehat{H}_{\Delta=0.5,\lambda=0}$; (c) $\widehat{H}_{\Delta=1,\lambda=0.4}$; (d) $\widehat{H}_{\Delta=1,\lambda=1}$ and (e) $\widehat{H}_{\Delta=0.5,\lambda=1}$. The solid line is the energy shell: Gaussian centered at $E_{\text{ini}}$ of width $\sigma_{\text{ini}}$ (see Tables \ref{table:initial}, \ref{table:ipr}); bin size = 0.05 $J$; $L=16$.}
\label{fig:shellNS}
\end{figure}

$|\rm{DW}\rangle$, $ |\rm{PS}\rangle$, and $ |\rm{NS}\rangle$ are chosen to magnify the effects of the anisotropy and of the NNN couplings.  Based on Tables~\ref{table:initial} and  \ref{table:ipr} and on Figs.~\ref{fig:shellPS} and \ref{fig:shellNS}, we analyze how the filling of the shell, $\text{IPR}_{\text{ini}}$ and $\sigma_ {\text{ini}}$ depend on $\Delta$, $\lambda$ and $L$. This is the first step for the understanding of the behavior of the fidelity, which is discussed in the following subsection. We find here that the best filling of the energy shell is by far associated with the N\'eel state under the strongly chaotic Hamiltonians.

As $\Delta$ increases, $E_{\text{ini}}$ is pushed to the edges of the spectrum, where the states get more localized. This dependence between $E_{\text{ini}}$ and $\Delta$ is seen in Table~\ref{table:initial} and in the panels for the energy distribution for $|\rm{PS}\rangle$ in Fig.~\ref{fig:shellPS} and for $|\rm{NS}\rangle$ in Fig.~\ref{fig:shellNS}   (distributions for $|\rm{DW}\rangle$ are shown in Refs.~\cite{Torres2014PRA89,Zangara2013}).  For $|\rm{DW}\rangle$ and  $|\rm{PS}\rangle$, $E_{\text{ini}}$ is further pushed to the borders as  $\lambda$  increases, whereas for  $|\rm{NS}\rangle$, $\lambda$ counterbalances the NN contributions and actually brings the state closer to the middle of the spectrum. Among the five cases in Fig.~\ref{fig:shellPS}, $E_{|\rm{PS}\rangle}$ is farthest from the center of the spectrum for the chaotic isotropic Hamiltonian [Fig.~\ref{fig:shellPS} (d)], while for the N\'eel state this happens for the integrable isotropic $\widehat{H}_F$ [Fig.~\ref{fig:shellNS} (a)]. Depending on $\lambda$, the energy of the initial state may also depend on $L$. In the integrable domain, a linear dependence on $L$ occurs for $|\rm{DW}\rangle$ and $ |\rm{NS}\rangle$, while in the chaotic regime with $\lambda=1$, it happens for $|\rm{DW}\rangle$ and $ |\rm{PS}\rangle$ (see Table~\ref{table:initial}).

The width of the energy shell does not depend on the anisotropy parameter, but it may be affected by the presence of NNN couplings. The shell  broadens with $\lambda$ for $|\rm{DW}\rangle$ and  $|\rm{PS}\rangle$ [Fig.~\ref{fig:shellPS}], but not for  $|\rm{NS}\rangle$ [Fig.~\ref{fig:shellNS}], since for this state $M_2=0$. The width is extensive on the system size for $|\rm{NS}\rangle$ ($M_1=L-1$) and $|\rm{PS}\rangle$ ($M_1=L/2, M_2=L-2$), but not for $|\rm{DW}\rangle$ ($M_1=1, M_2=2$). The energy shell for the domain wall has the smallest $\sigma_{\text{ini}}$ among the three initial states.  $|\rm{NS}\rangle$ has the largest $M_1$  and thus the largest $\sigma_{\text{ini}}$ when $\lambda =0$, but it is surpassed by $|\rm{PS}\rangle$ when $\lambda>1/\sqrt{2}$. 

Despite being, in general, close to Gaussian, the LDOS for the initial states considered differ with respect to the filling of the energy shell. The latter depends on the interplay between $\Delta$ and $\lambda$, and of course also on $L$. The values of the least square, used to quantify the deviation of the LDOS from the energy shell, are given in Table~\ref{table:ipr}. Small values indicate good filling of the shell. The exact values depend on the chosen bin size, which is somewhat arbitrary. It needs to be sufficiently small so that regions inside the shell where $|C_{\alpha}^{\text{ini}}|^2$ is very small can be detected. Our choice was made to guarantee that for the same $\sigma_{\text{ini}}$, the least square was smaller if $\text{IPR}_{\text{ini}}$ was larger. In Table~\ref{table:ipr} one finds also the values of $\text{IPR}_{\text{ini}}$, which give information about how much spread the initial states are in the energy eigenbasis. Since $\Delta$ pushes $E_{\text{ini}}$ to the edges of the spectrum, for a fixed $\sigma_{\text{ini}}$, better filling occurs for smaller anisotropy. This is confirmed with the values of least square in Table~\ref{table:ipr} and by comparing panels (a) with (b) and (d) with (e) in Figs.~\ref{fig:shellPS} and \ref{fig:shellNS}.  

The dependence of the filing on $\lambda$ is more subtle. For $|\rm{NS}\rangle$, since the NNN couplings simply push $E_{\text{ini}}$  to the center of the spectrum, the behavior is monotonic: the filling improves and $\text{IPR}_{|\text{NS}\rangle}$ increases with $\lambda$ be $\Delta$ equal to 0.5 or 1. For $|\rm{DW}\rangle$ and $|\rm{PS}\rangle$ this behavior holds only for $\Delta=0.5$. The chaotic anisotropic $\widehat{H}_{\Delta=0.5,\lambda=1}$ leads to the lowest least square value for the three $|{\text{ini}}\rangle$ [cf. Fig.~\ref{fig:shellPS} (e), Fig.~\ref{fig:shellNS} (e), and Table~\ref{table:ipr}], $|\rm{NS}\rangle$ being the most delocalized one. When $\Delta =1$ the improvement with $\lambda$ occurs only as the parameter goes from 0 to 0.4 and the connectivity increases, while from $\lambda=0.4$ to 1 the least square value increases. This happens because, despite the broadening of the LDOS, $\lambda$ pushes $E_{|\rm{DW}\rangle}$ and $E_{|\rm{PS}\rangle}$ to edge of the spectrum, where the states are more localized. This unfavorable combination causes the worst filling of the shell for $|\rm{PS}\rangle$ to occur for the {\em chaotic} isotropic Hamiltonian.  In this case the energy distribution  is skewed and spiky [Fig.~\ref{fig:shellPS} (d)] and $\text{IPR}_{|\text{PS}\rangle}$ has the lowest value [Table~\ref{table:ipr}]. 
 
\begin{table}
\caption{$E_\text{ini}$, $\sigma_\text{ini}$, least square (lsq) and $\text{IPR}_\text{ini}$ for $L = 16$.}
\begin{center}
\resizebox{8.5cm}{!}
{
\begin{tabular}{cccccl}
\hline \hline
&&\hspace{0.4 cm}$\text{E}_\text{ini}$&\hspace{0.4 cm}$\sigma_{\text{ini}}$&\hspace{0.4 cm}lsq&\hspace{0.4 cm}$\text{IPR}_\text{ini}$\\ 

&\hspace{0.3 cm}$\widehat{H}_{\Delta=1.0,\lambda=0.0}$&\hspace{0.4 cm}3.250&\hspace{0.4 cm}$0.50$ &\hspace{0.4 cm}$11.622$&\hspace{0.4 cm}$34.86$\\ 
 &\hspace{0.3 cm}$\widehat{H}_{\Delta=0.5,\lambda=0.0}$&\hspace{0.4 cm}1.625&\hspace{0.4 cm}$0.50$&\hspace{0.4 cm}$1.464$&\hspace{0.4 cm}$113.74$\\ 
$|\rm{DW}\rangle$ &\hspace{0.3 cm}$\widehat{H}_{\Delta=1.0,\lambda=0.4}$&\hspace{0.4 cm}4.250&\hspace{0.4 cm}$0.57$&\hspace{0.4 cm}$6.632$&\hspace{0.4 cm}$31.95$\\ 
 &\hspace{0.3 cm}$\widehat{H}_{\Delta=0.5,\lambda=0.4}$&\hspace{0.4 cm}2.125&\hspace{0.4 cm}$0.57$&\hspace{0.4 cm}$1.483$&\hspace{0.4 cm}$289.91$\\  
 &\hspace{0.3 cm}$\widehat{H}_{\Delta=1.0,\lambda=1.0}$&\hspace{0.4 cm}5.750&\hspace{0.4 cm}$0.87$&\hspace{0.4 cm}$11.335$&\hspace{0.4 cm}$28.69$\\ 
 &\hspace{0.3 cm}$\widehat{H}_{\Delta=0.5,\lambda=1.0}$&\hspace{0.4 cm}2.875&\hspace{0.4 cm}$0.87$&\hspace{0.4 cm}$1.047$&\hspace{0.4 cm}$368.14$\\ 
[0.2 cm]
\hline
&\hspace{0.3 cm}$\widehat{H}_{\Delta=1.0,\lambda=0.0}$&\hspace{0.4 cm}-0.250&\hspace{0.4 cm}$1.41$&\hspace{0.4 cm}$1.936$&\hspace{0.4 cm}$200.57$\\  
 &\hspace{0.3 cm}$\widehat{H}_{\Delta=0.5,\lambda=0.0}$&\hspace{0.4 cm}-0.125&\hspace{0.4 cm}$1.41$&\hspace{0.4 cm}$1.481$&\hspace{0.4 cm}$241.43$ \\ 
$ |\rm{PS}\rangle$ &\hspace{0.3 cm}$\widehat{H}_{\Delta=1.0,\lambda=0.4}$&\hspace{0.4 cm} -1.650&\hspace{0.4 cm}$1.60$&\hspace{0.4 cm}$0.938$&\hspace{0.4 cm}$592.72$\\
  &\hspace{0.3 cm}$\widehat{H}_{\Delta=0.5,\lambda=0.4}$&\hspace{0.4 cm}-0.825&\hspace{0.4 cm}$1.60$&\hspace{0.4 cm}$0.931$&\hspace{0.4 cm}$900.84$\\   
 &\hspace{0.3 cm}$\widehat{H}_{\Delta=1.0,\lambda=1.0}$&\hspace{0.4 cm}-3.750&\hspace{0.4 cm}$2.34$&\hspace{0.4 cm}$2.844$&\hspace{0.4 cm}$129.88$\\ 
 &\hspace{0.3 cm}$\widehat{H}_{\Delta=0.5,\lambda=1.0}$&\hspace{0.4 cm} -1.875&\hspace{0.4 cm}$2.34$&\hspace{0.4 cm}$0.543$&\hspace{0.4 cm}$586.56$\\
 \hline 
&\hspace{0.3 cm}$\widehat{H}_{\Delta=1.0,\lambda=0.0}$&\hspace{0.4 cm}-3.750&\hspace{0.4 cm}$1.94$&\hspace{0.4 cm}$4.609$&\hspace{0.4 cm}$72.15$\\ 
 &\hspace{0.3 cm}$\widehat{H}_{\Delta=0.5,\lambda=0.0}$&\hspace{0.4 cm}-1.875&\hspace{0.4 cm}$1.94$&\hspace{0.4 cm}$2.473$&\hspace{0.4 cm}$129.83$ \\ 
$ |\rm{NS}\rangle$ &\hspace{0.3 cm}$\widehat{H}_{\Delta=1.0,\lambda=0.4}$&\hspace{0.4 cm}-2.350&\hspace{0.4 cm}$1.94$&\hspace{0.4 cm}$1.105$&\hspace{0.4 cm}$336.78$\\ 
  &\hspace{0.3 cm}$\widehat{H}_{\Delta=0.5,\lambda=0.4}$&\hspace{0.4 cm}-1.175&\hspace{0.4 cm}$1.94$&\hspace{0.4 cm}$0.900$&\hspace{0.4 cm}$623.50$\\ 
 &\hspace{0.3 cm}$\widehat{H}_{\Delta=1.0,\lambda=1.0}$&\hspace{0.4 cm}-0.250&\hspace{0.4 cm}$1.94$&\hspace{0.4 cm}$0.237$&\hspace{0.4 cm}$1805.25$\\ 
 &\hspace{0.3 cm}$\widehat{H}_{\Delta=0.5,\lambda=1.0}$ &\hspace{0.4 cm}-0.125&\hspace{0.4 cm}$1.94$&\hspace{0.4 cm}$0.209$&\hspace{0.4 cm}$2071.92$\\ 
\hline
\hline
\end{tabular}
}
\end{center}
\label{table:ipr}
\end{table}

As a final remark, we note the unexpected relation between the width of the density of states [Table~\ref{table:DOS}] and that of the energy shell. For $L=16$, $\omega>\sigma_{\text{ini}}$ for $|\rm{DW}\rangle$ under the six Hamiltonians, but this is not the case when the initial state is $|\rm{PS}\rangle$ or $|\rm{NS}\rangle$. For the first, $\omega < \sigma_{\text{ini}}$ for $\widehat{H}_{\Delta=0.5,\lambda=1.0}$ and for the latter this happens for all Hamiltonians, except the strongly chaotic ones.

\subsection{Fidelity Decay}
\label{Sec:FidBasis}

\begin{figure}[htb]
\centering
\includegraphics*[width=4.5in]{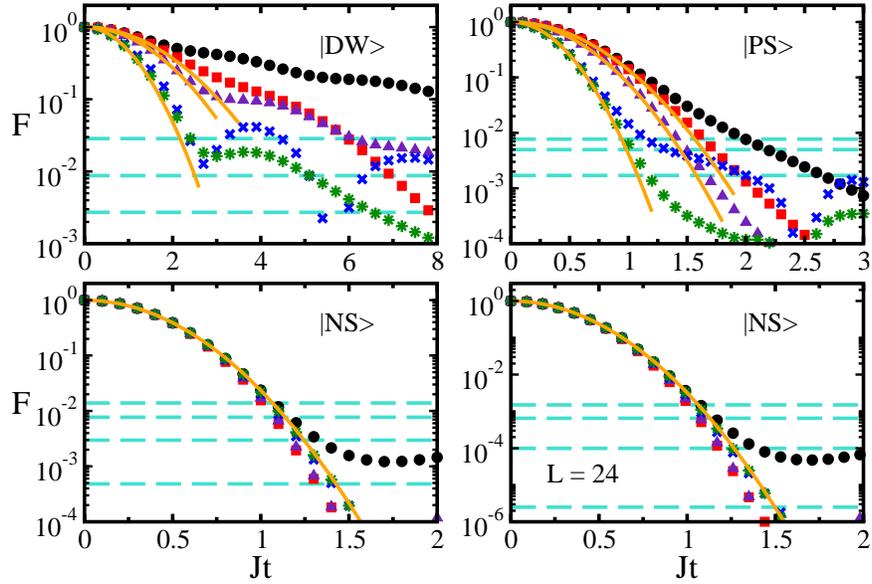}
\caption{(Color online)  Fidelity decay for the Hamiltonians: $\widehat{H}_{\Delta=1,\lambda=0}$ (circle),  $\widehat{H}_{\Delta=0.5,\lambda=0}$ (square), $\widehat{H}_{\Delta=1,\lambda=0.4}$ (triangle), $\widehat{H}_{\Delta=1,\lambda=1}$ (cross) and $\widehat{H}_{\Delta=0.5,\lambda=1}$ (star). The initial states are indicated; $L=16$ unless indicated otherwise.  Solid curves correspond to the analytical Gaussian expression in Eq. (\ref{eq:Fgauss}). The dashed horizontal lines give the saturation value $\text{IPR}_{\text{ini}}^{-1} $ (see Table~\ref{table:ipr}). For $|{\text{DW}} \rangle $ from top to bottom: $\widehat{H}_{\Delta=1,\lambda=0}$ (other isotropic cases are very close); $\widehat{H}_{\Delta=0.5,\lambda=0}$; $\widehat{H}_{\Delta=0.5,\lambda=1}$.  For $|{\text{PS}} \rangle $ from top to bottom: $\widehat{H}_{\Delta=1,\lambda=1}$; $\widehat{H}_{\Delta=1,\lambda=0}$ ($\widehat{H}_{\Delta=0.5,\lambda=0}$ is very close); $\widehat{H}_{\Delta=0.5,\lambda=1}$ ($\widehat{H}_{\Delta=1,\lambda=0.4}$ is very close). For $|{\text{NS}} \rangle $, both system sizes, from top to bottom: $\widehat{H}_{\Delta=1,\lambda=0}$;  $\widehat{H}_{\Delta=0.5,\lambda=0}$; $\widehat{H}_{\Delta=1,\lambda=0.4}$; $\widehat{H}_{\Delta=0.5,\lambda=1}$ ($\widehat{H}_{\Delta=1,\lambda=1}$ is very close). For $L=24$, the saturation values are obtained from an infinite time average with $Jt \in [1000,2000]$. All saturation values are larger than the one reached by a state evolved under GOE full random matrices, where $\text{IPR}_{\text{ini}}^{-1} \approx 3/{\cal D}$.}
\label{fig:fidelity02}
\end{figure}

The fidelity decay reflects the results of the energy distribution of $|{\text{ini}}\rangle $, as illustrated in Fig.~\ref{fig:fidelity02}. Overall, when the energy shell is well filled, the decay is Gaussian and this behavior may persist until saturation, as seen for the N\'eel state. In contrast, poor filling causes a mixture of Gaussian and exponential behavior, as shown for $|\rm{DW}\rangle$ and $|\rm{PS}\rangle$.

The fidelity decays slowly for $|\rm{DW}\rangle$, due to its low connectivity and narrow energy distribution. At short time the behavior is Gaussian and equal for systems with the same $\lambda$ (same $\sigma_{|\rm{DW}\rangle}$), but soon the curves for isotropic and anisotropic systems diverge, the first being slower than the latter, as expected from the filling of the shell. It is close to this point of separation that the exponential behavior takes over, although for the domain wall it does not remain until saturation. This state has a complicated dynamics at longer time, with the emergence of some plateaus indicating possible regions of pre-relaxation. Notice also that the infinite time average of the fidelity is very similar for the isotropic Hamiltonians, but the time for it to be reached depends on the strength of the NNN couplings, being shorter for larger $\lambda$. Among the Hamiltonians, $\widehat{H}_{\Delta=1,\lambda=1}$ leads to the fastest relaxation to equilibrium, since it combines largest $\sigma_{|\rm{DW}\rangle}$ and largest saturation value.

The fidelity decay for $|\rm{PS}\rangle$ shares common features with $|\rm{DW}\rangle$: at short time it is Gaussian and equal for Hamiltonians with the same $\lambda$, later it switches to an exponential behavior, and fastest relaxation to equilibrium happens for the chaotic isotropic system. However, contrary to $|\rm{DW}\rangle$, the exponential decay of $|\rm{PS}\rangle$ persists until close to equilibration and $\text{IPR}_{|\rm{PS}\rangle}^{-1} $ differs among the isotropic Hamiltonians. Furthermore, according to Table~\ref{table:initial}, the fidelity decay rate increases with $L$ for $|\rm{PS}\rangle$, whereas $\sigma_{|\rm{DW}\rangle}$ does not depend on the system size.

For the N\'eel state, where $\sigma_{|\rm{NS}\rangle}$ is dissociated from $\Delta $ and $\lambda$, the curves for $F(t)$ fall on top of each other for the five Hamiltonians considered in the figure. The decay is Gaussian until saturation. As mentioned before, this is in contrast with previous studies where the exponential behavior superseded (or was expected to supersede) the Gaussian decay before saturation \cite{Jalabert2001, Jacquod2001, Cerruti2002, Cucchietti2002, ProsenZnidaric2002, Prosen2002, Benenti2002,FlambaumARXIV,Flambaum2000A,Flambaum2001a,Flambaum2001b,Santos2012PRL,Santos2012PRE}. We note that the slight acceleration of $\widehat{H}_{\Delta=0.5,\lambda=0}$ and $\widehat{H}_{\Delta=1,\lambda=0.4}$ may be caused by the  presence of spikes far in the energy distribution. When these peaks are large and far in energy, they can add fast cosine decays to the Gaussian behavior.

The persistence of the Gaussian decay for $\sigma_{\text{ini}} t >1$ is not particular to the N\'eel state. We verified it for several site-basis vectors. In fact,  the majority of the site-basis vectors are much more delocalized than $|\rm{NS}\rangle$ and have comparable $\sigma_{\text{ini}}$. 

The N\'eel state emphasizes the role of the interplay between initial state and final Hamiltonian. It is impressive to find integrable and chaotic, isotropic and anisotropic systems, all leading to the same dynamics. The difference appears only at the saturation point. The infinite-time average value decreases monotonically from $\Delta=1$ to $\Delta=0.5$ and from $\lambda=0$ to $\lambda=1$ (see $\text{IPR}_{|\rm{NS}\rangle}$ in Table~\ref{table:ipr}). As a result $\widehat{H}_{\Delta=1,\lambda=0}$ is the first to reach equilibrium and $\widehat{H}_{\Delta=0.5,\lambda=1}$ is the last one. After relaxing, the fidelity fluctuates around $\text{IPR}_{|\rm{NS}\rangle}^{-1} $. The size of the fluctuations decrease exponentially with the system size [see Eq.~(\ref{Eq:sigmaF})].

An advantage of using site-basis vectors as initial states is the access that they give to exact analytical expressions for $E_{\text{ini}}$  and $\sigma_{\text{ini}}$. In general, one needs exact full diagonalization to find these values, which limits the system sizes that can be studied. For the dynamics, on the other hand, there are alternative methods, such as Krylov subspace techniques or density matrix renormalization group, that can deal with larger $L$. Having access to $\sigma_{\text{ini}}$ without the need to resort to exact diagonalization allows us to compare the analytical expression in Eq.~(\ref{eq:Fgauss}) with numerical results for $F(t)$ for $L>16$. In the bottom right panel of Fig.~\ref{fig:fidelity02}, we use EXPOKIT~\cite{Expokit,Sidje1998}  and confirm the Gaussian fidelity decay for the N\'eel state up to saturation also for $L=24$.

It remains to understand what exactly causes the transition from the Gaussian to the exponential behavior, when the shape of LDOS is approximately Gaussian. A rough estimate for this critical time was provided in \cite{Izrailev2006}. However, we found numerically a large range of values for this time. How it depends on $E_{\text{ini}}$,  $\sigma_{\text{ini}}$, $\text{IPR}_{\text{ini}}$ and other possible relevant factors is still unclear to us. We also find surprising that the fidelity decay for the N\'eel state under $\widehat{H}_{\Delta=1,\lambda=1}$ can be Gaussian until saturation, given that this state is very localized compared to others where the transition occurs. These questions are currently under investigation.

\section{Shannon Entropy}
\label{Sec:Shannon}

The results for the Shannon (information) entropy presented here reiterate those for the fidelity decay discussed in the previous section and anticipate the studies for observables in the next section. In particular, we show again that $\Delta$ does not affect the short time dynamics of initial states corresponding to site-basis vectors and that the N\'eel state does not depend on $\lambda$ either. However, the main goal of this section is to make known that  for site-basis vectors, including $|\rm{NS}\rangle$, the increase of the Shannon entropy transitions from quadratic to linear, but the latter cannot be reproduced by analytical expressions derived in~\cite{Flambaum2001b}. These derivations need to be reassessed.

The Shannon entropy is a delocalization measure that, just like IPR, depends on the basis. Written in the eigenstates of the final Hamiltonian, the Shannon entropy for a certain $|\text{ini}\rangle$ corresponds to the diagonal entropy \cite{Polkovnikov2011,Santos2011PRL}, which, after the quench, is a static quantity. In contrast, if the Shannon entropy is written in the eigenstates of $\widehat{H}_I$, it will evolve in time.  In this case, it quantifies the gradual spreading of the initial state and increased participation of the basis vectors in time.

The evolution of the Shannon entropy in the site-basis vectors $| n \rangle$ is given by
\begin{equation}
\text{Sh}(t) = - \sum_n W_n(t) \ln W_n(t) , 
\label{eq.sh1}
\end{equation}
where
\begin{equation}
W_n(t) =\left| \left\langle n \left| e^{-i \widehat{H}_{F} t} \right| \text{ini} \right\rangle \right|^2 \:,
\end{equation}
and $W_{\text{ini}}(t) = F(t)$. 

\begin{figure}[h]
\centering
\includegraphics*[width=4.25in]{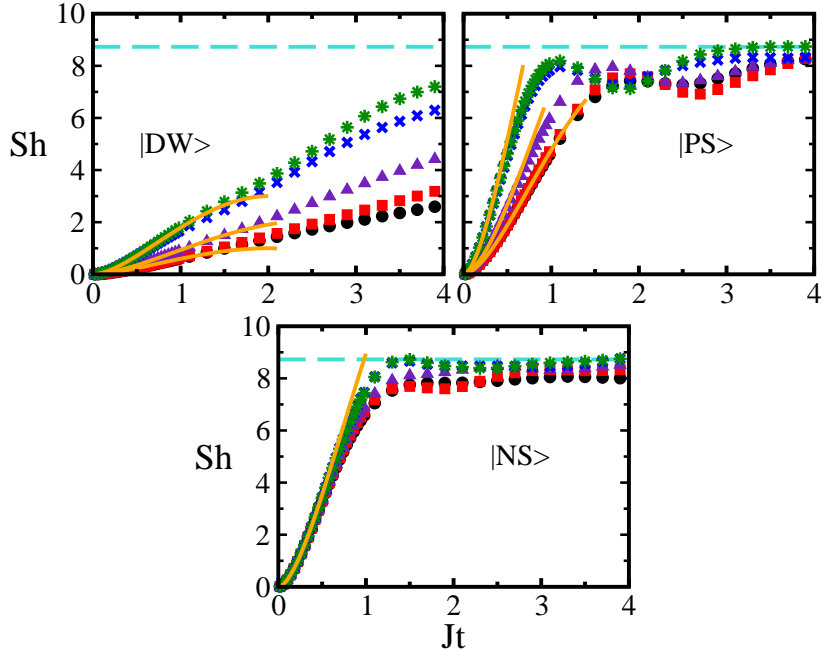}
\caption{(Color online) Shannon entropy in the site-basis vectors for the final Hamiltonians: $\widehat{H}_{\Delta=1,\lambda=0}$ (circle),  $\widehat{H}_{\Delta=0.5,\lambda=0}$ (square), $\widehat{H}_{\Delta=1,\lambda=0.4}$ (triangle), $\widehat{H}_{\Delta=1,\lambda=1}$ (cross) and $\widehat{H}_{\Delta=0.5,\lambda=1}$ (star). The initial states are indicated; $L=16$. Solid curves correspond to the approximation in Eq.~(\ref{eq.sh2}). The dashed horizontal line indicates the value of the Shannon entropy reached by full random matrices from GOEs, $\text{Sh}_{\text{GOE}} \sim \ln(0.48 \cal{D})$ \cite{ZelevinskyRep1996, Izrailev1990}.  }
\label{fig:shannon}
\end{figure} 

Obtaining an analytical expression for $\text{Sh}(t)$ is non-trivial due to the dependence on the overlap between the evolved initial state and the other basis vectors. But we can estimate the short-time dynamics by expanding Eq.~\eqref{eq.sh1} and using the expression for fidelity $F(t)$ given in Eq.~\eqref{eq:Fgauss}. We obtain
\begin{eqnarray}
&&\text{Sh}(t) \approx \sigma_{\text{ini}}^2  t^2 \!-\! t^2 \! \sum_{n \neq \text{ini}}  \left| \left\langle n \left| \widehat{H}_{F} \right| \text{ini} \right\rangle \right|^2  \!\ln \!\left[ t^2  \left| \left\langle n \left| \widehat{H}_{F} \right| \text{ini} \right\rangle \right|^2 \right] 
\nonumber \\ 
&&
= \sigma_{\text{ini}}^2  t^2 - \frac{J^2 t^2}{4} \left[ M_1 \ln \left( \frac{J^2 t^2}{4} \right) + \lambda^2 M_2 \ln \left( \frac{\lambda^2 J^2 t^2}{4} \right) \right] .
\label{eq.sh2}
\end{eqnarray}

From the equation above, it is clear that $\Delta$ does not affect the evolution of $\text{Sh} $ when $ \sigma_{\text{ini}} t < 1$. States evolving under isotropic or anisotropic Hamiltonians show very similar behavior, as seen in  Fig.~\ref{fig:shannon} for $|\text{ini}\rangle =|\rm{DW}\rangle, |\rm{PS}\rangle$, and $|\rm{NS}\rangle$. The role of the anisotropy becomes noticeable at longer times, particularly for the domain wall, where the evolution is slow and gives enough time for a clear separation of the curves before saturation.  

In terms of regime, the entropy grows faster in the chaotic domain for $|\rm{DW}\rangle$ and $|\rm{PS}\rangle$. In contrast, the initial evolution for the N\'eel state does not depend on $\lambda$, since $M_2=0$. The expression for the entropy simplifies to 
\begin{equation}
\text{Sh}_{|\text{NS} \rangle } (t) \approx \frac{J^2 t^2 (L-1)}{4} \left[1 - \ln \left( \frac{J^2 t^2}{4} \right) \right] ,
\label{eq:ShNS}
\end{equation}
showing dependence only on the system size. The curves for $|\rm{NS}\rangle$ in Fig.~\ref{fig:shannon} coincide.

For short times, Fig.~\ref{fig:shannon} indicates good agreement between the numerical results and the approximated Eq.~(\ref{eq.sh2}). 
At later times, there is a visible transition from a quadratic to a linear behavior before saturation. 
This linear increase of the entropy is, however, not well described by the expressions obtained in Ref.~\cite{Flambaum2001b} for initial states with a Breit-Wigner energy distribution and used also in \cite{Santos2012PRL,Santos2012PRE} for initial states with a Gaussian LDOS. 
\begin{figure}[htb]
\centering
\includegraphics*[width=4.in]{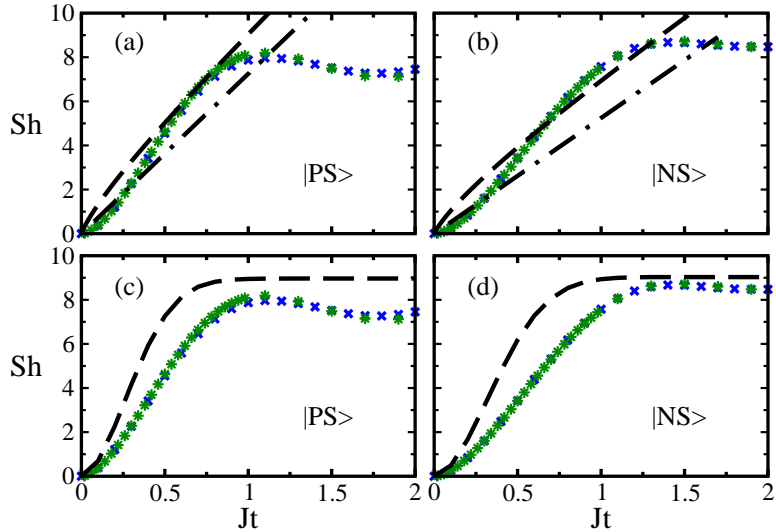}
\caption{(Color online) Comparison between the numerical results for the Shannon entropy with Eq.~(\ref{eq:Slinear}) (dashed line) and Eq.~(\ref{eq:SlinearSimple}) (dot-dashed line) in panels (a) and (b), and with Eq.~(\ref{eq:Sanalytical}) (dashed line) in panels (c) and (d); $\widehat{H}_{\Delta=1,\lambda=1}$ (cross) and $\widehat{H}_{\Delta=0.5,\lambda=1}$ (star); $L =16$. For Eq.~(\ref{eq:Sanalytical}), $N_{pc}$ was obtained from the average in $Jt \in  [3000,4000]$. The initial states are indicated.}
\label{fig:shannonProblem}
\end{figure}

The derivation of the analytical expression in \cite{Flambaum2001b} is based on a cascade model that separates sets of states $|n\rangle$ in classes. Each class is successively populated by states directly coupled with those from the previous class, leading to
\begin{equation}
\text{Sh} (t)= \Gamma_{\text{ini}} t \ln M 
+ \Gamma_{\text{ini}} t - e^{-\Gamma_{\text{ini}} t} \sum_{j=0}^{\infty} \frac{(\Gamma_{\text{ini}} t)^j}{j!} \ln \frac{(\Gamma_{\text{ini}} t)^j}{j!} ,
\label{eq:Slinear}
\end{equation}
where $\Gamma_{\text{ini}}$ is the width of the Breit-Wigner. In Refs.~\cite{Flambaum2001b,Santos2012PRL,Santos2012PRE}, the connectivity of the initial state was large, so the last two terms were smaller than the first one. The increase of the entropy was then well captured by
\begin{equation}
\text{Sh} (t)= \Gamma_{\text{ini}} t \ln M .
\label{eq:SlinearSimple}
\end{equation}
In Refs.~\cite{Santos2012PRL,Santos2012PRE}, $\Gamma_{\text{ini}}$ was substituted by the width $\sigma_{\text{ini}}$ of the Gaussian LDOS. A semi-analytical expression was also proposed to describe the dynamics at both short and long times~\cite{Flambaum2001b}. It corresponds to
\begin{equation}
\text{Sh}(t) = - F(t) \ln F(t) -
\left[ 1- F(t)\right] \ln \left( \frac{1- F(t)}{N_{pc}}\right),
\label{eq:Sanalytical}
\end{equation}
where $N_{pc} = \overline{\exp (\text{Sh})}$  is obtained numerically by performing an average in a large time interval after relaxation. 

None of the three expressions can describe the results in Fig.~\ref{fig:shannon}. The problem is caused by the low connectivity of the initial states, where $M \propto L$. The above expressions are valid for initial states with large connectivity, $M \propto {\cal D}$. They were obtained for initial states corresponding to good mean-field basis states. In this latter case, the cascade model assumption of negligible probabilities of return from one class to previous classes holds, while it is violated for our $|\text{ini}\rangle$'s. 

In  Fig.~\ref{fig:shannonProblem} we compare our results for the Shannon entropy with Eqs.(\ref{eq:Slinear}), (\ref{eq:SlinearSimple}), and (\ref{eq:Sanalytical}). For the semi-analytical expression (\ref{eq:Sanalytical}), we calculated $F(t) $ and $N_{pc}$ numerically.  The illustrations are for $|\rm{PS}\rangle$ [Fig.~\ref{fig:shannonProblem} (a) and (c)] and $|\rm{NS}\rangle$ [Fig.~\ref{fig:shannonProblem} (b) and (d)] evolving under  $\widehat{H}_{\Delta=0.5,\lambda=1}$ and  $\widehat{H}_{\Delta=1,\lambda=1}$. In this case, $|\rm{PS}\rangle$ has the largest connectivity among the states studied, $M=3L/2-2$,  and $|\rm{NS}\rangle$ is the state that  best fills the energy shell. The disagreement between the numerical results and the three expressions is evident. 

One cannot discard, however, the possibility of extending the cascade model to include also low-connectivity-states. For this, it will be necessary to take into account that the connectivities may not be approximately constant for each class, especially for the first classes. For site-basis vectors, for example, we find great discrepancies. The N\'eel state is directly coupled to $L-1$ states, while these states couple with $L^2-5$ states. These two connectivities are very different and much smaller than ${\cal D}$. 

\section{Few-Body Observables}
\label{Sec:Obs}

The analysis of the evolution of few-body observables is, of course, more involved than the study of the fidelity decay. It depends on the overlaps between the evolved $|\text{ini}\rangle$ and other basis vectors, as in the Shannon entropy, and also on the details of the observables $\widehat{A}$. However, a simple general picture, valid at short times, can be constructed for observables that commute with $\widehat{H}_I$. In this case, the fidelity, and therefore $\sigma_{\text{ini}}$, plays an important role in $A (t)$.

The evolution of the observables is given by
\begin{eqnarray}
A(t) &=& F(t) A(0)  \nonumber \\
&+& \sum_{n\neq \text{ini}} \langle \text{ini} | e^{i \widehat{H}_F t} | \text{ini} \rangle A_{\text{ini} ,n} \langle n | e^{- i \widehat{H}_F t} | \text{ini} \rangle + \sum_{n\neq \text{ini}} \langle \text{ini} | e^{i \widehat{H}_F t} | n \rangle A_{n,\text{ini} } \langle \text{ini} | e^{- i \widehat{H}_F t} | \text{ini} \rangle \nonumber \\
&+&\sum_{n,m\neq \text{ini}} \langle \text{ini} | e^{i \widehat{H}_F t} | n \rangle A_{n, m} \langle m | e^{- i \widehat{H}_F t} | \text{ini} \rangle ,
\label{eq:TotObs}
\end{eqnarray}
where $A_{n, m} = \langle  n | \widehat{A} | m\rangle$ and $|n\rangle $ are the eigenstates of $\widehat{H}_I$.
When $[\widehat{H}_I,\widehat{A}]=0$, since $|\text{ini}\rangle$ is one of the eigenstates of $\widehat{H}_I$, $A_{\text{ini} ,n} =0$ for $|n\rangle \neq |\text{ini}\rangle$ and the second line in Eq.~(\ref{eq:TotObs}) cancels. The dominant terms of the expansion in time give simply
\begin{equation}
A(t) \approx \left( 1-\sigma_{\text{ini}}^2 t^2 \right)  A(0) + t^2 \sum_{n \neq \text{ini}} | \langle n | \widehat{H}_F  | \text{ini} \rangle |^2 A_{n, n}.
\label{eq:ObsCommute}
\end{equation}

Summary of the results presented below:

$\bullet$ In Secs.~\ref{sec:mag}, \ref{sec:corre}, \ref{sec:struct}, and \ref{Sec:current}, we explore the case of initial states corresponding to site-basis vectors, where $ \widehat{H}_I$ is the Ising part of the Hamiltonian~(\ref{ham}). For these states, $\langle n | \widehat{H}_F  | \text{ini} \rangle$, and $\sigma_{\text{ini}}$ do not depend on the anisotropy parameter, so the short-time evolutions generated by isotropic and anisotropic final Hamiltonians are equivalent. For the N\'eel state, not even $\lambda$ is important, so integrable and chaotic Hamiltonians also lead to a very similar initial relaxation of the observables.

$\bullet$ Sections~\ref{sec:mag}, \ref{sec:corre}, and \ref{sec:struct} deal with three observables that commute with $ \widehat{H}_I$: two that are local in space, the local magnetization and the spin-spin correlation in the $z$ direction, and one that is nonlocal in space, namely the structure factor in the $z$ direction. As expected from equation \eqref{eq:ObsCommute}, their short-time evolution is quadratic in time. For the structure factor the dynamics depends on momentum and system size. Section~\ref{Sec:current} discusses the spin current, which does not commute with $ \widehat{H}_I$, so Eq.~(\ref{eq:ObsCommute}) cannot be used. We find that the short-time dynamics of this observable, even though linear in time, shows a dependence on $\Delta$ and $\lambda$ that is comparable to what is seen for the other three observables.

$\bullet$ In Sec.~\ref{sec:entang}, we show that we can construct specific initial states, where Eq.~(\ref{eq:ObsCommute}) can still closely dictate the short-time dynamics of observables that do not commute with $ \widehat{H}_I$.

\subsection{Local magnetization}
\label{sec:mag}

The on-site magnetization, $\widehat{S}_j^z$, is a simple and yet useful observable frequently measured experimentally~\cite{JurcevicARXIV}. 
It is straightforward to show that for $|\rm{DW}\rangle$, $|\rm{PS}\rangle$, and $|\rm{NS}\rangle$ the magnetization in the middle of the chain behaves, at short times, as
\begin{eqnarray}
S_{L/2}^{z,|\rm{DW}\rangle}(t)  &=&  S^z_{L/2}(0)  \left[ 1 - \frac{J^2 t^2}{2} (1 + \lambda ^2) \right], \nonumber \\
S_{L/2}^{z,|\rm{PS}\rangle}(t)  &=&  S^z_{L/2}(0)  \left[ 1 - \frac{J^2 t^2}{2} (1 +2 \lambda ^2) \right], \nonumber \\
S_{L/2}^{z,|\rm{NS}\rangle}(t)  &=&  S^z_{L/2}(0)  \left[1 -  J^2 t^2 \right].  \nonumber
\end{eqnarray}
In the integrable regime, $S_{L/2}^{z,|\rm{NS}\rangle}(t)$ changes faster than for the other two states, since for $|\rm{NS}\rangle$, the excitation on site $L/2$ has two neighboring sites to hop to, while for $|\rm{DW}\rangle$ and $|\rm{PS}\rangle$ it has only one site. When directly couplings between NNN are included and $\lambda>1/\sqrt{2}$, $S_{L/2}^{z,|\rm{PS}\rangle}(t)$ becomes the fastest to evolve, since the excitation has now three sites to hop to, while $|\rm{NS}\rangle$ and $|\rm{DW}\rangle$  have only two. At the edges of the chain, border effects slow down the dynamics. 

The above approximations agree well with our numerical results at short times (not shown). The curves with the same value of $\lambda$ coincide for $|\rm{DW}\rangle$ and $|\rm{PS}\rangle$, whereas for $|\rm{NS}\rangle$, the curves fall on top of each other, independently of $\Delta$ or $\lambda$.

\subsection{Spin-spin correlations}
\label{sec:corre}

The spin-spin correlation between sites $i$ and $j$  is given by 
\begin{equation}
\widehat{C}^{\mu}_{i,j} (t)=  \widehat{S}^\mu_{i} \widehat{S}^\mu_{j} ,\quad \mu =x,z.
\label{cxx} 
\end{equation}
In the $z$ direction and using Eq.~(\ref{eq:ObsCommute}) and Table~\ref{table:initial}, we find for neighboring sites in the middle of the chain that
\begin{eqnarray}
&&C^{z,|\rm{DW}\rangle}_{\frac{L}{2},\frac{L}{2}+1}(t)  =  C^{z,|\rm{DW}\rangle}_{\frac{L}{2},\frac{L}{2}+1}(0)   \left[ 1 - \frac{J^2 \lambda ^2 t^2}{2}  \right], \label{eq:Cdw} \\
&&C^{z,|\rm{PS}\rangle}_{\frac{L}{2},\frac{L}{2}+1}(t)   =  C^{z,|\rm{PS}\rangle}_{\frac{L}{2},\frac{L}{2}+1}(0)  \left[ 1 - J^2 t^2 (1 +2 \lambda ^2) \right], \label{eq:Cps} \\
&&C^{z,|\rm{NS}\rangle}_{\frac{L}{2},\frac{L}{2}+1}(t)  =  C^{z,|\rm{NS}\rangle}_{\frac{L}{2},\frac{L}{2}+1}(0)  \left[1 -  J^2 t^2 \right].  \label{eq:Cns}
\end{eqnarray}
The expression for $|\rm{PS}\rangle$ above holds for $\text{mod}(L,4)=0$, when the spins on sites $L/2,L/2+1$ are parallel. 
When $\text{mod}(L,4) \neq 0$, and the two middle spins are anti-parallel, the expression changes to  $C^{z,|\rm{PS}\rangle}_{\frac{L}{2},\frac{L}{2}+1}(0)  [ 1 - J^2 t^2 (1 +3 \lambda ^2)/2 ]$.

\begin{figure}[htb]
\centering
\includegraphics*[width=4.5in]{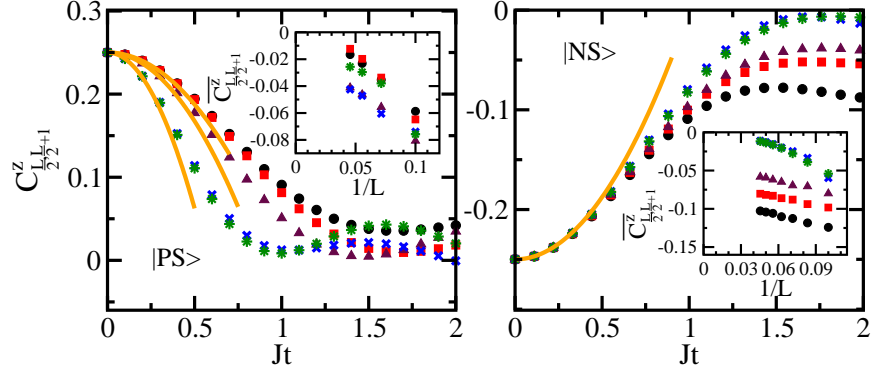}
\caption{(Color online) Spin-spin correlation in the $z$ direction between neighboring sites in the middle of the chain.
The final Hamiltonians are:  $\widehat{H}_{\Delta=1,\lambda=0}$ (circle),  $\widehat{H}_{\Delta=0.5,\lambda=0}$ (square), $\widehat{H}_{\Delta=1,\lambda=0.4}$ (triangle), $\widehat{H}_{\Delta=1,\lambda=1}$ (cross) and $\widehat{H}_{\Delta=0.5,\lambda=1}$ (star). The initial states are indicated. Main panels: $L=16$. Solid curves are the analytical results from Eq.~(\ref{eq:Cps}) and (\ref{eq:Cns}). The insets
show the scaling with $L$ of the infinite time average of $C^{z}_{\frac{L}{2},\frac{L}{2}+1}$, computed in $Jt \in [3000,4000]$.}
\label{fig:czzns}
\end{figure}

As expected and confirmed numerically (not shown), the magnitude of $C^{z,|\rm{DW}\rangle}_{\frac{L}{2},\frac{L}{2}+1} (t)$ decays slowly, especially in the integrable domain. Figure~\ref{fig:czzns} compares the longitudinal correlation for $|\rm{PS}\rangle$ and $|\rm{NS}\rangle$ evolving under the same final Hamiltonians. Similarly to what was seen for fidelity, the initial decay of  the magnitude of $C^{z}_{\frac{L}{2},\frac{L}{2}+1} (t)$ for the N\'eel state is independent of the regime of $\widehat{H}_F$, while for $|\rm{PS}\rangle$ it is faster in the chaotic domain. These distinct behaviors, anticipated from Eqs.~(\ref{eq:Cps}) and (\ref{eq:Cns}), emphasize the significance of the initial state also for the dynamics of observables.

After a long time, $C^{z}_{\frac{L}{2},\frac{L}{2}+1}(t)$ fluctuates around the equilibrium value $\overline{C^{z}}_{\frac{L}{2},\frac{L}{2}+1}$, the fluctuations decreasing exponentially with system size~\cite{Zangara2013}. The saturation value  is closest to zero when $E_{\text{ini}}$ is closest to the center of the spectrum. This happens for $|\rm{PS}\rangle$ with the integrable Hamiltonians and for $|\rm{NS}\rangle$ with the strongly chaotic Hamiltonians, as can be seen in the insets of Fig.~\ref{fig:czzns}.

The insets of Fig.~\ref{fig:czzns} give the scaling of $\overline{C^{z}}_{\frac{L}{2},\frac{L}{2}+1}$ with system size. For $|\rm{NS}\rangle$, the correlations for the strongly chaotic Hamiltonians approach zero as $L$ increases, while the results indicate that integrable and weakly chaotic Hamiltonians may retain memory in the thermodynamic limit. However, we cannot discard the possibility of an acceleration towards zero for $L$'s larger than the ones considered here. The results for $|\rm{PS}\rangle$ are less conclusive. For this state, the direction of the spins in the middle of the chain depend on $L$. This causes the saturation value for small $L$ to oscillate significantly from $\text{mod}(L,4) \neq 0$ to $\text{mod}(L,4) = 0$. To try to delineate a pattern, we show only the results for $\text{mod}(L,4) \neq 0$. The correlations decrease with system size, but more points are necessary for an extrapolation to the thermodynamic limit.

\subsection{Structure Factor}
\label{sec:struct}

The structure factor is the Fourier transform of the spin-spin correlations, being therefore a nonlocal observable in space. In the $z$ direction,
\begin{equation}
\widehat{s}_f^{z}(\kappa ) = \frac{1}{L} \; \sum_{j,k=1}^L e^{-i \kappa (j-k)} \;
\widehat{S}_j^{z} \; \widehat{S}_k^{z} \;
=\frac{1}{4} + \frac{2}{L} \; \sum_{u=1}^{L-1} \; \cos \left(  \kappa u \right) \; \sum_{v=1}^{L-u} \; \widehat{S}_v^{z} \; \widehat{S}_{v+u}^{z} .
\end{equation}
Above, $ \kappa=2\pi p/L$ stands for momentum and $p=0,1,2 \ldots, L$ is a positive integer.

\begin{figure}[htb]
\centering
\includegraphics*[width=3.75in]{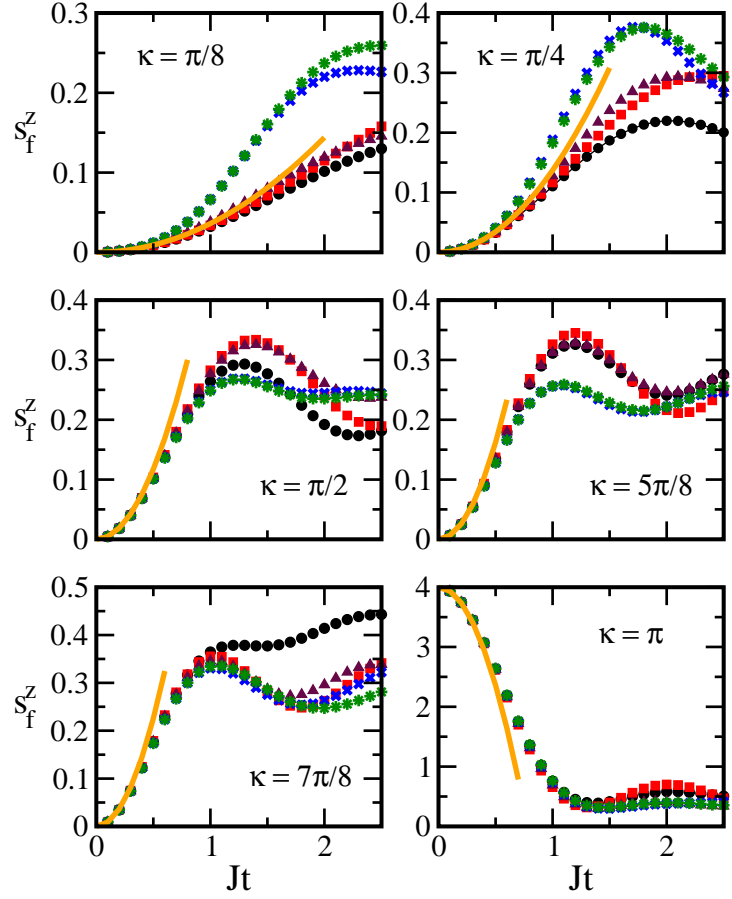}
\caption{(Color online) 
Structure factor in the $z$ direction for the N\'eel state. The final Hamiltonians are:  $\widehat{H}_{\Delta=1,\lambda=0}$ (circle),  $\widehat{H}_{\Delta=0.5,\lambda=0}$ (square), $\widehat{H}_{\Delta=1,\lambda=0.4}$ (triangle), $\widehat{H}_{\Delta=1,\lambda=1}$ (cross) and $\widehat{H}_{\Delta=0.5,\lambda=1}$ (star). The values of $\kappa$ are indicated in the panels. Solid curves are the analytical results from Eq.~(\ref{eq:SFsimple}); $L=16$.}
\label{sfzns}
\end{figure}

The evolution of the structure factor depends on $\kappa$. For example, for the N\'eel state, Eq.~(\ref{eq:ObsCommute}) becomes
\begin{equation}
s_f^{z,|\rm{NS}\rangle}(\kappa,t) \approx (1 - \sigma_{|\rm{NS}\rangle}^2 t^2) s_f^{z,|\rm{NS}\rangle}(\kappa,0) 
+\frac{J^2 t^2}{4} (L-1) \beta (\kappa) 
\label{eq:SFsimple}
\end{equation}  
where
\begin{eqnarray}
&&s_f^{z,|\rm{NS}\rangle}(\kappa,0) = \frac{1}{4} + \frac{1}{2L} \sum_{u=1}^{L-1} \cos (\kappa u) (-1)^u (L-u) , \nonumber \\
&& \beta(\kappa) =  \frac{1}{4} + \frac{1}{2L}\left[ - \cos \kappa + \sum_{u=1}^{L-2} \cos (\kappa u) (-1)^u (L-u-4) \right] \nonumber ,
\end{eqnarray}
The magnitude of the term ${\cal O}(t^2)$ is largest when $\kappa=\pi$ (for $L=16$, it is $-6.56 J^2 t^2$). It decreases abruptly for $p=L/2-1$ (for $L=16$, it is $0.90 J^2 t^2$) and then gradually until $p=1$ (for $L=16$, it is $0.04 J^2 t^2$). It is only for $\kappa=\pi$ that $s_f^{z,|\rm{NS}\rangle}(\kappa,0) \neq 0$, which allows for a large contribution from the width of the energy shell. This dependence on $\kappa$ is depicted in Fig.~\ref{sfzns}. 
Since the magnitude of ${\cal O}(t^2)$ becomes very small as $p\rightarrow 1$, higher order terms, with effects from $\Delta$ and $\lambda$, become significant already at short times, which explains the early divergence of the curves for different Hamiltonians. In contrast, for $\kappa=\pi$ the curves coincide up to the vicinity of the saturation point.

Contrary to the local observables $C^{z}_{i,j} (t)$ and $S^z_j(t)$, the evolution of the structure factor depends also on $L$. The decay is faster as $L$ increases, although how fast it is depends again on $\kappa$.
When $\kappa=\pi$, Eq.~(\ref{eq:SFsimple}) simplifies to
\begin{equation}
s_f^{z,|\rm{NS}\rangle}(\pi,t) \approx \frac{L}{4} - \frac{J^2 t^2}{2} \left( \frac{2}{L} - 3 +L \right),
\end{equation}
which shows a linear dependence of the term ${\cal O}(t^2)$  on $L$. For other $\kappa$'s the dependence on $L$ is much smaller.

A similar analysis can be extended to $|\rm{DW}\rangle$ and $|\rm{PS}\rangle$. In general, for these states, the evolution of $s_f^{z,|\rm{ini}\rangle}(\kappa,t)$ at short times depends on $\lambda$, as seen previously for other observables. The dependence on $\kappa$ is again present, although it is not the same found for the N\'eel state. For instance, for $|\rm{PS}\rangle$, the effects of $\sigma_{|\rm{PS}\rangle}$ occur only when $\kappa=\pi/2$, where $s_f^{z,|\rm{PS}\rangle}(\kappa,0) \neq 0$.
Interestingly, when $\kappa=\pi$, $s_f^{z,|\rm{PS}\rangle}(\kappa,t)$ shows no dependence on $\lambda$ at short times. This happens because $s_f^{z,|\rm{PS}\rangle}(\kappa,0)$ and the contributions from the states directly coupled with $|\rm{PS}\rangle$ via NNN couplings are zero when $\text{mod}(L,4) =0$ and when $\text{mod}(L,4) \neq 0$ they cancel each other.

\subsection{Local spin current}
\label{Sec:current}

The spin current is an observable of great interest for studies about quantum transport~\cite{Zotos1997,Zotos1999,Heidrich2003,Zotos2005,Heidrich2007,Karrasch2013}, which motivates having a closer look at it. Even though its evolution cannot be cast in the form of Eq.~(\ref{eq:ObsCommute}), the short-time dynamics shows again a simple dependence on $\lambda$ that is comparable to what was found for the previous observables. In particular, the behavior for $|\rm{NS}\rangle$ is again very similar for all Hamiltonians considered.

The local spin current, $I_{s,j}$, is associated with the conservation of total spin in the $z$ direction, ${\cal S}^z$, and obeys the continuity equation~\cite{Santos2011},
\[
\frac{\partial S^z_j}{\partial t} + div(I_{s,j})=0.
\]
In the bulk, the local spin current agrees with the result from a periodic chain~\cite{Zotos1997},
\[
-i [H, S_j^z] = div(I_{s,i})= (I_{s,j} - I_{s,j-1}),
\]
which leads to
\begin{equation}
I_{s,j} = J(S_j^x S_{j+1}^y - S_j^y S_{j+1}^x)  
+\lambda J (S_j^x S_{j+2}^y - S_j^y S_{j+2}^x
+S_{j-1}^x S_{j+1}^y - S_{j-1}^y S_{j+1}^x).
\end{equation}

\begin{figure}[htb]
\centering
\includegraphics*[width=4.75in]{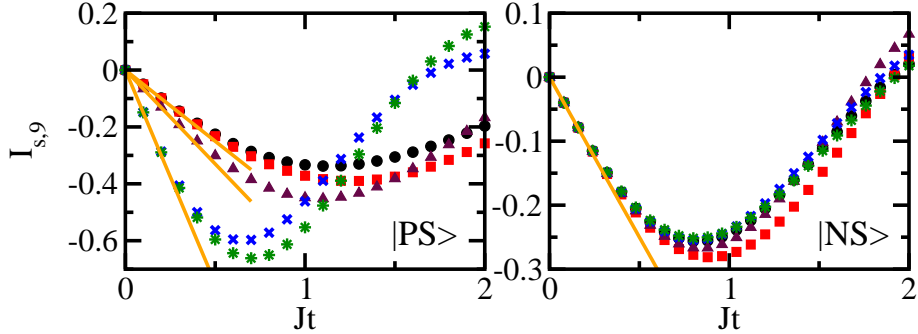}
\caption{(Color online)  Local spin current $I_{s,9}$ for the final Hamiltonians:  $\widehat{H}_{\Delta=1,\lambda=0}$ (circle),  $\widehat{H}_{\Delta=0.5,\lambda=0}$ (square), $\widehat{H}_{\Delta=1,\lambda=0.4}$ (triangle), $\widehat{H}_{\Delta=1,\lambda=1}$ (cross) and $\widehat{H}_{\Delta=0.5,\lambda=1}$ (star). The initial states are indicated; $L=16$.}
\label{fig:current}
\end{figure}

This observable does not commute with $\widehat{H}_I$ and $I_{s,j} (0)=0$. Also in contrast with the previous observables, $\langle n | I_{s,j} | \text{ini} \rangle$ is imaginary, so the dominant term in the expansion of Eq.~(\ref{eq:TotObs}) is
\[
I_{s,j}(t) \approx - i t \sum_{n\neq \text{ini}}  \langle \text{ini} | I_{s,j} | n \rangle \langle n | \widehat{H}_F | \text{ini} \rangle +
i t \sum_{n\neq \text{ini}}  \langle n | I_{s,j} | \text{ini} \rangle \langle \text{ini}| \widehat{H}_F|n\rangle .
\]
If the pair of spins on sites $(j,j+1)$ are parallel, as happens for $|\rm{PS}\rangle$ when $\text{mod}(L,4)=0$, then $ \langle n | I_{s,j} | \text{ini} \rangle =0$. This explains why, in Fig.~\ref{fig:current}, where $L=16$, we selected $I_{s,9}$. It is straightforward to show that for pairs of anti-parallel spins on sites $(j,j+1)$,
\begin{eqnarray}
&& I_{s,j}^{|\rm{PS}\rangle}(t) = \frac{-J^2 t}{2} - \lambda^2 J^2 t ,  \\
&& I_{s,j}^{|\rm{NS}\rangle}(t) = \frac{-J^2 t}{2} . 
\label{eq:I_NS}
\end{eqnarray}
The above expressions capture well the initial dynamics of the local spin current shown in Fig.~\ref{fig:current}. They make evident the lack of any influence of $\Delta$ at short times, the dependence on $\lambda$ for $|\rm{PS}\rangle$, and its absence for $|\rm{NS}\rangle$. For the N\'eel state, the proximity of the curves for very different Hamiltonians and even beyond the range of validity of Eq.~(\ref{eq:I_NS}) is again remarkable.

\subsection{Entangled State}
\label{sec:entang}

The simple picture developed for the short-time dynamics of $\widehat{C}^{z}_{i,j}$ becomes less trivial when dealing with $\widehat{C}^{x}_{i,j}$. One reason is the disappearance of the role of the fidelity (and $\sigma_{\text{ini}}$), since $C^{x}_{i,j}(0)=0$ when the initial state is a site-basis vector. The other issue is the contributions, already at ${\cal O}(t^2)$, from terms such as $\langle n | \widehat{H}^2_F  | \text{ini} \rangle$, which bring into play the effects of the anisotropy. Nevertheless, one can construct particular initial states, where behaviors close to that of Eq.~(\ref{eq:ObsCommute}) can be recovered also for $\widehat{C}^{x}_{i,j}$, and similar observables.

Here, this is illustrated for an initial state corresponding to an entangled state, $|\rm{ES}\rangle$ which contains an EPR pair on sites $L/2$ and $L/2+1$, 
\[
|\rm{ES}\rangle = \frac{1}{\sqrt{2}} \left\{ | \cdot \uparrow \downarrow \cdot \rangle + | \cdot \downarrow \uparrow \cdot \rangle \right\},
\]
where,
\[
| \cdot \uparrow \downarrow \cdot \rangle = | \uparrow \uparrow \uparrow \uparrow \ldots \downarrow \downarrow \downarrow  \ldots \uparrow \downarrow   \ldots \downarrow \downarrow \downarrow \downarrow  \ldots  \uparrow \uparrow \uparrow \rangle
\]
and 
\[
| \cdot \downarrow \uparrow \cdot \rangle =  | \uparrow \uparrow \uparrow \uparrow \ldots \downarrow \downarrow \downarrow \ldots \downarrow \uparrow \ldots \downarrow \downarrow \downarrow \downarrow \ldots \uparrow \uparrow \uparrow \rangle .
\]
On each side of the EPR pair, there is a domain wall. 
To simplify the analysis, we fix system sizes such that $\text{mod}(L,4)=0$. 

The energy and width of the energy distribution of $|\rm{ES}\rangle$ are
\begin{eqnarray}
&&E_{|\rm{ES}\rangle}  =\frac{J}{ 2} + \frac{ J\Delta}{ 4} [(L-9) + (L-14)\lambda] , \nonumber \\
&& \sigma_{|\rm{ES}\rangle} = \frac{ J}{ 2} \sqrt{3+2\lambda+6\lambda^2}. \nonumber
\end{eqnarray}
Their dependence on $\Delta$, $\lambda$, and $L$ is similar to that for $|\rm{DW}\rangle$, apart from an additional term $\propto \lambda$ in $\sigma_{|\rm{ES}\rangle}^2$. The energy shell for $|\rm{ES}\rangle$ is, however,  much better filled (compare Table~\ref{table:ES} and Table~\ref{table:ipr}). For the entangled state with $\widehat{H}_{\Delta=0.5,\lambda=1}$, the least square is even better than that for the N\'eel state.

\begin{figure}[htb]
\centering
\includegraphics*[width=4.5in]{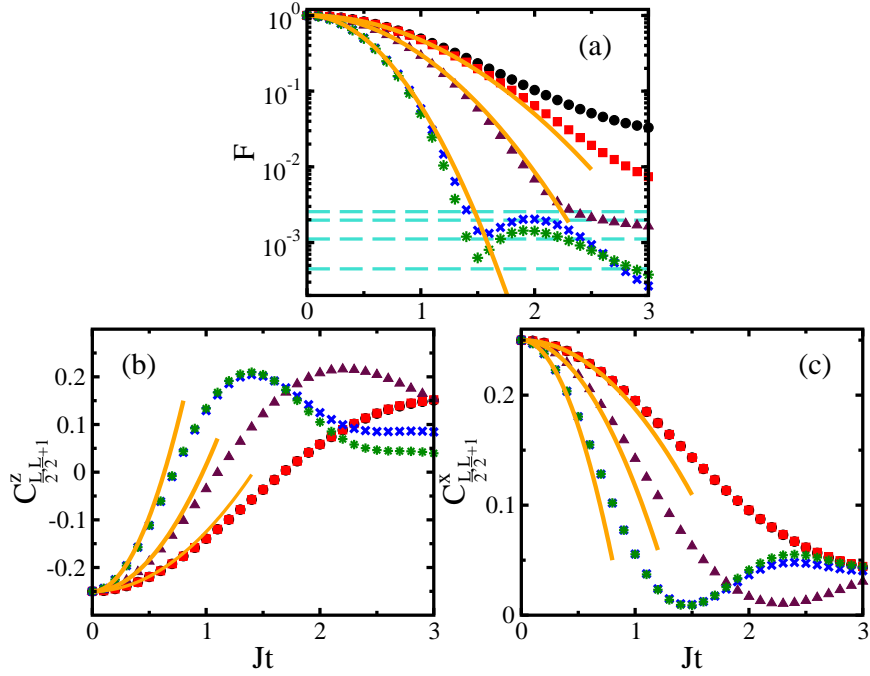}
\caption{(Color online) Time evolution of the fidelity (a) and spin-spin correlations in the longitudinal (b) and transversal (c) directions for $|\rm{ES}\rangle$; $L=16$. The final Hamiltonians are:  $\widehat{H}_{\Delta=1,\lambda=0}$ (circle),  $\widehat{H}_{\Delta=0.5,\lambda=0}$ (square), $\widehat{H}_{\Delta=1,\lambda=0.4}$ (triangle), $\widehat{H}_{\Delta=1,\lambda=1}$ (cross) and $\widehat{H}_{\Delta=0.5,\lambda=1}$ (star). Solid curves are analytical results given by Eqs.~(\ref{eq:Fgauss}), (\ref{eq:CzES}) and (\ref{eq:CxES}) with parameters given in Table~\ref{table:ES}. In panel (a), dashed horizontal lines are the saturation values for the fidelity. From top to bottom: $\widehat{H}_{\Delta=1,\lambda=0}$; $\widehat{H}_{\Delta=1,\lambda=0.4}$; $\widehat{H}_{\Delta=0.5,\lambda=0}$ ($\widehat{H}_{\Delta=1,\lambda=1}$ is very close); $\widehat{H}_{\Delta=0.5,\lambda=1}$.}
\label{fig:ES}
\end{figure}

The fidelity decay of $|\rm{ES}\rangle$ is shown in Fig.~\ref{fig:ES} (a). The decay is slower than that of $|\rm{PS}\rangle$ and $|\rm{NS}\rangle$ (cf. Fig.~\ref{fig:fidelity02}) due to the domain wall structure that leads to smaller values of $\sigma_{|\rm{ES}\rangle}$. Yet, in the chaotic domain, the behavior is Gaussian until very close to saturation, as anticipated from the low values of least square.

\begin{table}
\caption{$E_\text{ini}$, $\sigma_\text{ini}$, least square (lsq) and $\text{IPR}_\text{ini}$ for $L = 16$.}
\begin{center}
\resizebox{8.5cm}{!}
{
\begin{tabular}{cccccl}
\hline \hline
& &\hspace{0.4 cm}$E_\text{ini}$&\hspace{0.4 cm}$\sigma_\text{ini}$&\hspace{0.4 cm}lsq& \hspace{0.4 cm}$\text{IPR}_\text{ini}$\\ [0.1 cm]
&\hspace{0.3 cm}$\widehat{H}_{\Delta=1.0,\lambda=0.0}$&\hspace{0.4 cm}$2.250$&\hspace{0.4 cm}$ 0.87$&\hspace{0.4 cm}$1.764$&\hspace{0.4 cm}$391.15$\\ 
 &\hspace{0.3 cm}$\widehat{H}_{\Delta=0.5,\lambda=0.0}$ &\hspace{0.4 cm}$1.375$&\hspace{0.4 cm}$0.87$&\hspace{0.4 cm}$0.746$&\hspace{0.4 cm}$822.90$\\ 
$|\rm{ES}\rangle$ &\hspace{0.3 cm}$\widehat{H}_{\Delta=1.0,\lambda=0.4}$ &\hspace{0.4 cm}$2.450$&\hspace{0.4 cm}$1.09$&\hspace{0.4 cm}$0.603$&\hspace{0.4 cm}$506.76$\\ 
  &\hspace{0.3 cm}$\widehat{H}_{\Delta=0.5,\lambda=0.4}$ &\hspace{0.4 cm}$1.475$&\hspace{0.4 cm}$1.09$&\hspace{0.4 cm}$0.163$&\hspace{0.4 cm}$1601.97$\\ 
 &\hspace{0.3 cm}$\widehat{H}_{\Delta=1.0,\lambda=1.0}$ &\hspace{0.4 cm}$2.750$&\hspace{0.4 cm}$1.66$&\hspace{0.4 cm}$0.427$&\hspace{0.4 cm}$898.23$\\ 
 &\hspace{0.3 cm}$\widehat{H}_{\Delta=0.5,\lambda=1.0}$&\hspace{0.4 cm}$1.625$&\hspace{0.4 cm}$1.66$&\hspace{0.4 cm}$0.162$&\hspace{0.4 cm}$2223.89$\\ [0.2 cm]
\hline
\hline
\end{tabular}
}
\end{center}
\label{table:ES}
\end{table}

Using the approximation $ \langle \cdot \uparrow \downarrow \cdot  \mid e^{-i \widehat{H}_F t}  \mid \text{ES} \rangle  \approx  \langle \cdot \downarrow \uparrow \cdot \mid e^{-i \widehat{H}_F t} \mid \text{ES} \rangle  $, which holds at short times, we can expand $C_{\frac{L}{2},\frac{L}{2}+1}^z(t)$ to ${\cal O}(t^2)$ as in Eq.~(\ref{eq:ObsCommute}). This gives
\begin{equation}
C_{\frac{L}{2},\frac{L}{2}+1}^z(t) = C_{\frac{L}{2},\frac{L}{2}+1}^z(0) \left[ 1 - \frac{J^2 t^2}{2} (1 + 2 \lambda + 2 \lambda^2) \right].
\label{eq:CzES}
\end{equation}

For $|\rm{ES}\rangle$, since $|\cdot \uparrow \downarrow \cdot \rangle$ is directly coupled with $| \cdot \downarrow \uparrow \cdot \rangle$, $\widehat{C}^{x}_{\frac{L}{2},\frac{L}{2}+1}(0) \neq 0$ and we can find an expansion similar to that in Eq.~(\ref{eq:ObsCommute}) also for the correlation in the $x$ direction. It gives
\begin{equation}
C_{\frac{L}{2},\frac{L}{2}+1}^x(t) = C_{\frac{L}{2},\frac{L}{2}+1}^x(0) \left[ 1 - \frac{J^2 t^2}{4} (1 + 2 \lambda + 2 \lambda^2) \right].
\label{eq:CxES}
\end{equation}
At short times, both expressions above show good agreement with the numerical results in Figs.~\ref{fig:ES} (b) and (c), respectively. Since there is no dependence on $\Delta$, the curves for equal $\lambda$ are very similar. In fact, they practically coincide until close to saturation. 

\section{Conclusion}
\label{Sec:Summary}

We studied isolated interacting quantum systems quenched far from equilibrium. The initial state $|\text{ini}\rangle$ was an eigenstate of an initial Hamiltonian $\widehat{H}_I$ and it evolved according to a final Hamiltonian $\widehat{H}_F$. We analyzed numerically and analytically the behavior in time of fidelity, Shannon entropy in the basis of $\widehat{H}_I$, and few-body observables. The focus was on initial states, system models, and observables that are accessible to current experiments with optical lattices. We showed that the system dynamics depends not only on the initial state or on the final Hamiltonian, but on the interplay between the two. Depending on the initial state,  different $\widehat{H}_F$'s may lead to very similar evolutions.

The fidelity is the Fourier transform of the energy distribution of the initial state. This distribution was referred to here as LDOS.  For quenches that lead to equivalent LDOS, the fidelity decay for different Hamiltonians is comparable. We investigated single-peaked LDOS. They are limited by the shape of the density of states of $\widehat{H}_F$. When the final Hamiltonians are full random matrices, the LDOS is semicircular, which leads to the lower bound for the fidelity decay: $F(t) =  [{\cal J}_1( 2 \sigma_{\text{ini}} t)]^2/(\sigma_{\text{ini}}^2 t^2)$, where ${\cal J}_1$ is the Bessel function of first kind and $\sigma_{\text{ini}}$ is the standard deviation of the LDOS. If instead of many-body interactions, as implied by full random matrices, only two-body interactions are considered, then the maximum LDOS is Gaussian and, consequently, the fastest fidelity decay is also Gaussian: $F(t) = \exp (- \sigma_{\text{ini}}^2  t^2)$.  The Gaussian behavior can persist up to saturation. In this latter case, the relaxation time is  $t_R=\sqrt{\ln(\text{IPR}_{\text{ini}}) }/\sigma_{\text{ini}}$, where $\text{IPR}_{\text{ini}}$ measures the level of delocalization of the initial state in the energy eigenbasis.  

The fidelity decay in systems with two-body interactions is directly related with the strength of the perturbation. For initial states away from the borders of the spectrum, we find three main cases: (i) For very small perturbations, the LDOS is close to a delta function and the dynamics is extremely slow. (ii) In the intermediate perturbation regime, the LDOS becomes Lorentzian and the fidelity decay is exponential. (iii) In the strong perturbation regime, the LDOS finally reaches a Gaussian shape, reflecting the form of the density of states of these systems, and the fidelity behavior is also Gaussian.

For initial states corresponding to site-basis vectors, we derived an expression that captures the evolution of the Shannon entropy at very short times. General analytical and semi-analytical expressions that had been successfully employed in the past~\cite{Flambaum2001b,Santos2012PRL,Santos2012PRE} did not match our results. To reassess these expressions, one will need to take into account the low connectivity of initial states that correspond to site-basis vectors.

The analysis of few-body observables $\widehat A$ was performed for local magnetization, spin-spin correlations, structure factor, and local spin current. Initial states corresponding to site-basis vectors are eigenstates of the first three observables in the $z$ direction, that is $[\widehat{H}_I,\widehat A]=0$. The short-time dynamics of such observables is quadratic in time and, for $A(0)\neq 0$, $\sigma_{\text{ini}} $ plays a central role in their evolution. The local spin current is not part of this general picture and its short-time evolution is linear.

We showed that the N\'eel state, which is a key state in magnetism and can be prepared in optical lattices, has a very interesting behavior. Since for this state, $\sigma_{\text{ini}} =J\sqrt{L-1}/2$, the fidelity decay and the short-time dynamics of the Shannon entropy and of the few-body observables where $[\widehat{H}_I,\widehat A]=0$ do not depend on the values of the anisotropy parameter or on the regimes of the final Hamiltonians. The evolution is very similar, be the final Hamiltonian integrable, chaotic, isotropic, or anisotropic.

Our results for fidelity and observables may contribute to the establishment of a general description for isolated quantum systems far from equilibrium. The analysis of the relationship between LDOS and fidelity decay can shed light on recent studies about fidelity and Loschmidt echo in the context of quench dynamics~\cite{Genway2010,Shchadilova2014,Andraschko2014,SchiroARXIV}. It is also likely to benefit the development of methods for quantum control and studies about the statistics of work done on quenched systems. In connection with the latter, we stress that the LDOS corresponds to the probability distribution of the work needed to take quantum systems out of equilibrium.


\begin{acknowledgments}
This work was supported by the  NSF grant No.~DMR-1147430. E.J.T.H. acknowledges partial support from CONACyT, Mexico.
\end{acknowledgments}


\end{document}